\documentclass[10pt,preprint]{aastex}

\newcommand{\deltaa}{{\ensuremath{\Delta\alpha}}}
\newcommand{\lbreak}{{\ensuremath{\lambda_{break}}}}
\newcommand{\cosi}{{\ensuremath{\cos i}}}
\newcommand{\alphav}{{\ensuremath{\alpha_{visco}}}}
\newcommand{\irasobj}{IRAS~F07546+3928}

\newcommand{\Lx}{{\ensuremath{L_{X}}}}
\newcommand{\Lxfuv}{{\ensuremath{L_{XFUV}}}}
\newcommand{\Lfuv}{{\ensuremath{L_{FUV}}}}
\newcommand{\Lfuvo}{{\ensuremath{L_{FUVO}}}}
\newcommand{\Lir}{{\ensuremath{L_{IR}}}}
\newcommand{\Lbol}{{\ensuremath{L_{Bol}}}}

\newcommand{\msun}{M_\odot}
\newcommand{\msunyr}{\ensuremath{M_\odot\,{\rm yr}^{-1}}}
\newcommand{\mbh}{\ensuremath{M_{BH}}}
\newcommand{\mdot}{\ensuremath{\dot{M}}}

\newcommand{\lledd}{\ensuremath{L/L_{Edd}}}
\newcommand{\lbol}{\ensuremath{L_{Bol}}}

\newcommand{\lllambda}{\ensuremath{\lambda L_{\lambda}}}
\newcommand{\flambda}{\ensuremath{f_{\lambda}}}
\newcommand{\fnu}{\ensuremath{f_{\nu}}}

\newcommand{\nhi}{$N_{\rm HI}$}

\newcommand{\ax}{{\ensuremath{\alpha_{\rm x}}}}

\newcommand{\alphal}{{\ensuremath{\alpha_{\lambda}}}}

\newcommand{\alambda}{\ensuremath{\alpha_{\lambda}}}
\newcommand{\anu}{\ensuremath{\alpha_{\nu}}}
\newcommand{\afuv}{\ensuremath{\alpha_{FUV}}}
\newcommand{\auvoa}{\ensuremath{\alpha_{{UVO}}}}
\newcommand{\auvob}{\ensuremath{\alpha_{{Ored}}}}
\newcommand{\aored}{\ensuremath{\alpha_{{Ored}}}}
\newcommand{\auvo}{\ensuremath{\alpha_{{UVO}}}}

\newcommand{\ovi}{O\,{\sc vi}}
\newcommand{\ovil}{O\,{\sc vi}\,$\lambda$1034}
\newcommand{\lya}{\ensuremath{{\rm Ly}{\alpha}}}
\newcommand{\lyal}{Ly{\sc $\alpha$}\,$\lambda$1216}
\newcommand{\lybl}{Ly{\sc $\beta$}\,$\lambda$1026}
\newcommand{\lygl}{Ly{\sc $\gamma$}\,$\lambda$973}

\newcommand{\oil}{O\,{\sc i}\,$\lambda$1304}

\newcommand{\hb}{\ensuremath{{\rm H}{\beta}}}

\newcommand{\siiv}{Si\,{\sc iv}}

\newcommand{\heii}{He\,{\sc ii}}

\newcommand{\heiiol}{He\,{\sc ii}\,$\lambda$4686}

\newcommand{\oiii}{[O\,{\sc iii}]}
\newcommand{\oiiil}{[O\,{\sc iii}]\,$\lambda$5007}
\newcommand{\oiiill}{[O\,{\sc iii}]\,$\lambda\lambda$4959,5007}

\newcommand{\feii}{Fe\,{\sc ii}}

\newcommand{\hst}{{\it HST}}
\newcommand{\fuse}{{\it FUSE}}
\newcommand{\ebv}{\ensuremath{\mbox{E(B-V)}}}

\newcommand{\kms}{\ensuremath{{\rm km~s}^{-1}}}
\newcommand{\kev}{\ensuremath{\mbox{keV}}}
\newcommand{\ergs}{\ensuremath{\mbox{ergs s}^{-1}}}

\newcommand{\fluxl}{erg s$^{-1}$ cm$^{-2}$ \AA$^{-1}$}

\newcommand{\Hoeq}{\ensuremath{H_0=75\,\kms \mbox{Mpc}^{-1}}}
\shorttitle{Quasars and the Big Blue Bump}
\shortauthors{Shang et al.}

\begin{document}

%\title{AGN SEDs and Black Hole Accretion}
\title{QUASARS AND THE BIG BLUE BUMP}


\author{Zhaohui Shang,\altaffilmark{1}
Michael S. Brotherton,\altaffilmark{1}
Richard F. Green,\altaffilmark{2}
Gerard A. Kriss,\altaffilmark{3}
\\
Jennifer Scott,\altaffilmark{3}
Jessica Kim Quijano,\altaffilmark{3}
Omer Blaes,\altaffilmark{4}
Ivan Hubeny,\altaffilmark{5}
%Tal Alexander,\altaffilmark{6}
John Hutchings,\altaffilmark{6}
\\
Mary Elizabeth Kaiser,\altaffilmark{7}
Anuradha Koratkar,\altaffilmark{8}
William Oegerle,\altaffilmark{9}
and Wei Zheng\altaffilmark{7}
}

\altaffiltext{1}{Department of Physics and Astronomy, University of Wyoming,
Laramie, WY 82071, USA; shang@uwyo.edu, mbrother@uwyo.edu}
\altaffiltext{2}{Kitt Peak National Observatory, National Optical 
Astronomy Observatories, P.O.Box 26732, 950 North Cherry
Avenue, Tucson, AZ 85726, USA}
\altaffiltext{3}{Space Telescope Science Institute, 3700 San
Martin Drive, Baltimore, MD 21218, USA}
\altaffiltext{4}{Department of Physics, University of California, Santa
Barbara, CA 93106, USA}
\altaffiltext{5}{National Optical Astronomy Observatories, 
P.O.Box 26732, 950 North Cherry Avenue, Tucson, AZ 85726, USA}
%\altaffiltext{6}{Weizmann Institute of Science, P.O.Box 26,
%Rehovot 76100, Israel}
\altaffiltext{6}{Herzberg Institute of Astrophysics, National Research
Council Canada, Victoria, BC V9E~2E7, Canada}
\altaffiltext{7}{Center for Astrophysical Sciences, Department
of Physics and Astronomy, The John Hopkins University,
Baltimore, MD 21218, USA}
\altaffiltext{8}{Goddard Earth Sciences and Technology Center, 3.002
South Campus, University of Maryland Baltimore County, 1000 Hilltop
Circle, Baltimore, MD 21250, USA}
\altaffiltext{9}{NASA Goddard Space Flight Center, 
Code 681, Greenbelt, MD 20771, USA}

\begin{abstract}

We investigate the ultraviolet-to-optical spectral energy
distributions (SEDs) of 17 active galactic nuclei (AGNs) using
quasi-simultaneous spectrophotometry spanning 900--9000\AA\ (rest
frame). We employ data from the Far Ultraviolet Spectroscopic Explorer
(\fuse), the Hubble Space Telescope (\hst), and the 2.1-meter
telescope at Kitt Peak National Observatory (KPNO).  Taking advantage
of the short-wavelength coverage, we are able to study the so-called
``big blue bump,'' the region where the energy output peaks, in
detail.  Most objects exhibit a
spectral break around 1100\AA.  Although this result is formally
associated with large uncertainty for some objects, there is strong
evidence in the data that the far-ultraviolet spectral region is below
the extrapolation of the near-ultraviolet-optical slope, indicating a
spectral break around 1100\AA.
  We compare the behavior of our sample to those of non-LTE
thin-disk models covering a range in black-hole mass, Eddington ratio,
disk inclination, and other parameters. The distribution of ultraviolet-optical
spectral indices redward of the break, and far-ultraviolet indices
shortward of the break, are in rough agreement with the models.
However, we do not see a correlation between the far-ultraviolet
spectral index and the black hole mass, as seen in some accretion
disk models.  We argue that the observed spectral break is intrinsic
to AGNs, although intrinsic reddening as well as Comptonization can
strongly affect the far-ultraviolet spectral index.  We make our data
available online in digital format.

\end{abstract}




\section{INTRODUCTION}

The spectral energy distribution (SED) of AGNs contains a significant
feature in the ultraviolet (UV) to optical region, known as ``the
big blue bump.''  This feature is thought to be thermal emission from 
an optically thick accretion disk feeding a massive black hole 
\citep[e.g.,][]{Shie78,MalSar82}, and it has been argued that its energy peak 
lies in the unobserved extreme ultraviolet (EUV, $\sim$100--912\AA) region 
\citep[e.g.,][]{MF87}.  To determine the shape and the
peak of the big blue bump is of particular importance, since this
provides critical information on the structure and condition of
the inner-most region in AGNs as well as on the ionizing flux that
powers the emission lines.

\citet{Zhen97} constructed composite AGN spectra from \hst\
spectra, and found that the UV-optical power-law continuum breaks
at around 1000\AA.  This was later confirmed in a similar \hst\
composite using a larger sample \citep{Telf02}.  \citet{Laor97}, also
using composite spectra, found that the soft X-ray continuum matches
up with the extrapolation of the \hst\ composite, consistent with
the idea that the UV bump actually peaks in far-ultraviolet (FUV,
912--2000\AA), rather
than the EUV.  However, such a spectral break is not seen in the
recent \fuse\ composite spectra for low-redshift
AGNs \citep{Scot04}, although some individual sources do show a spectral break.  
\citet{Scot04} noted that this \fuse\
sample has a median luminosity one order of magnitude lower
than that of the \hst\ sample \citep{Zhen97}, and it is possible
the break wavelength is luminosity dependent.  
% It is important to
% realize that composite spectra may be misleading, and the 
% spectra of individual objects must be used to check the results of
% the composite spectra, and to test continuum models of all sorts.
It is important to realize that composite spectra represent a complex
average of many individual spectra. Examination of the individual
spectra in a composite may lead to greater insights, and they can be
used to understand the overall characteristics of a composite in a
more physical way. 
For instance, the spectrum of 3C273 shows a break near the Lyman limit
\citep{Kris99}, and reasonable fits to the spectrum with accretion
disk models have been reported \citep{Kris99, Blae01}.

It is well accepted that AGNs are powered by accretion onto 
massive black holes.  Many accretion disk models have
been built to predict the AGN continuum and compare with observed
spectra  \citep[e.g.,][]{SunMal89,LaoNet89,Laor90,CzeZby91,Hube00,Hube01}.
Since the AGN local environment is not well known, and the models are
still relatively very simple, it is not possible for the models to
predict the detailed features in the observed spectra.  However,
large-scale features in the AGN SEDs can be reproduced by disk
models.  Simple models produce a large continuum discontinuity
at the Lyman limit, but it is not often seen in real objects
\citep[e.g.,][]{Kora92}.  Such a 
feature could be smeared out by relativistic smearing and Comptonization
\citep[e.g,][]{HsuBla98,Kris99,Blae01,Hube01}.  This results in a bump or a
spectral break, instead of an edge, in the vicinity of Lyman limit,
which resembles the spectral breaks seen in some observed AGN spectra.
\citet{Hube00} constructed geometrically thin accretion disk models
with non-LTE atmospheres, including a full treatment of general
relativistic effects in the disk structure.  We choose this specific
model and compare the spectral properties in the FUV-optical region
for a sample of 17 low-redshift AGNs with the model expectations.

We characterize the spectral continuum with broken power-laws.
Unless noted, the power-law indices we use through this paper are all
$\alambda$ ($\flambda \propto \lambda^{\alambda}$)
except that the soft X-ray spectral index \ax\ is \anu\ 
($\fnu \propto \nu^{\anu}$).
It is easy to convert between $\alambda$
and $\anu$  ($\alambda+\anu=-2$).  
For cosmology, we choose $\Lambda=0$,
\Hoeq, and $q_0=0.5$.
Since the objects in this sample are at low redshifts, the
results in the paper are not affected by the choice of cosmology.

%%%%%%%%%%%%%%%%%%%%%%%%%%%%%%%%%%%%%%%%%%%%%%%%%%%%%%%%%%%%%
\section{SAMPLE AND DATA}

The \fuse\ AGN program \citep{Kris00} has surveyed more than 100
of the UV-brightest AGNs, of which about 20 were also observed
in an \hst\ spectral snapshot survey during 1999--2000.  The \fuse\
observations were scheduled as close in time as possible with the
\hst\ snapshot observations.  Ground-based optical spectra were also
obtained during the same period at KPNO.  We excluded a few objects
because of the lack of an optical spectrum (NGC~3783, low declination),
strong host galaxy contamination (NGC~3516), or strong variability
(NGC~5548, also no simultaneous \hst\ spectrum).  As a result,
we have compiled a sample of 17 AGNs, with quasi-simultaneous
spectrophotometry covering rest wavelength from 900--9000\AA.
This is a heterogeneous sample with low redshift ($z<0.5$).

\begin{deluxetable}{llcccccccclc}
%\tabletypesize{\scriptsize}
\tabletypesize{\tiny}
\rotate
\tablecaption{Sample and Observation Log\label{tb:log}}
\tablewidth{0pt}
\tablehead{
\colhead{} & \colhead{} & \colhead{} & \colhead{} & 
\multicolumn{5}{c}{Observation Date}  &
\multicolumn{2}{c}{Dataset ID} &
\colhead{\it CalFUSE} \\ \cline{5-8} \cline{10-11}
\colhead{Object} & \colhead{Other Name} & 
\colhead{$z$\tablenotemark{a}} & 
\colhead{\ebv\tablenotemark{b}} & \colhead{\fuse} &
\colhead{\hst} & \colhead{Opt. Blue} & \colhead{Opt. Red} & &
\colhead{\fuse\ ID} & \colhead{\hst\ ID} &
\colhead{Version} 
}
\startdata
%Object	&	Other Name	&	z&	E(B-V)&	FUSE date	&	\hst\ date	&	Blue date	&	Red date	&	&	 FUSE ID	&	\hst\ FUV		\hst\ NUV	&	CalFuse	\\
3C273	&	PG1226+023	&	0.1576	&	0.021	&	2000-04-23	&	2000-03-16	&	2000-02-25	&	2000-02-26	&	&	 P1013501	&	O5G045JZQ	,	O5G045K0Q	&	v2.0.5	\\
3C351	&	PG1704+608	&	0.3730	&	0.023	&	1999-10-17	&	1992-02-15	&	1999-10-09	&	1990-09-20	&	&	 Q1060101	&	Y0VM0103T	,	Y0RV0G04T\tablenotemark{c}	&	v2.2.2	\\
4C+34.47	&	B2 1721+34	&	0.2055	&	0.037	&	2000-06-09	&	2000-06-25	&	1999-10-09	&	2000-02-27	&	&	 P1073501	&	O5G077Y2Q	,	O5G077Y3Q	&	v2.2.2	\\
IRAS F07546+3928	&	MS 0754.6+3928	&	0.0953	&	0.066	&	2002-02-11	&	2000-01-29	&	1999-10-09	&	1999-10-08	&	&	 S6011801	&	O5G018LPQ	,	O5G018LQQ	&	v2.1.7	\\
MRK290	&	PG1534+580	&	0.0303	&	0.015	&	2000-03-16	&	2000-06-02	&	2000-02-25	&	2000-02-27	&	&	 P1072901	&	O5G070TLQ	,	O5G070TMQ	&	v1.8.7	\\
MRK304	&	PG2214+139	&	0.0657	&	0.073	&	2000-07-16	&	2000-06-19	&	1999-10-07	&	1999-10-08	&	&	 P1073901	&	O5G082ANQ	,	O5G082AOQ	&	v2.0.5	\\
MRK506	&		&	0.0428	&	0.031	&	2000-06-08	&	2000-06-24	&	1999-10-09	&	2000-02-27	&	&	 P1073401	&	O5G076UEQ	,	O5G076UFQ	&	v2.0.5	\\
MRK509	&		&	0.0345	&	0.057	&
1999-11-06	&	1992-06-21	&	1999-12-11	&
1999-12-11	&	&	X0170102\tablenotemark{d}	&	Y0YA0302T	,	Y0YA0305T\tablenotemark{e}	&	v2.1.7	\\
NGC3516	&		&	0.0883	&	0.042	&	2000-04-17	&	2000-04-20	&		&		&	&	 P1110404	&	O5G032T7Q	,	O5G032T8Q	&	v2.1.7	\\
NGC3783	&		&	0.0097	&	0.119	&	2000-02-02	&	2000-05-17	&		&		&	&	 P1013301	&	O5G039LBQ	,	O5G039LCQ	&	v1.8.7	\\
PG0052+251	&		&	0.1544	&	0.047	&	1999-10-03	&	1999-10-01	&	1999-10-07	&	1999-10-08	&	&	 P1070101	&	O5G003NAQ	,	O5G003NBQ	&	v2.0.5	\\
PG0947+396	&		&	0.2057	&	0.019	&	2001-01-06	&	2000-06-15	&	1999-10-09	&	2000-02-27	&	&	 A0600101	&	O5G023NPQ	,	O5G023NQQ	&	v2.2.2	\\
PG0953+414	&		&	0.2338	&	0.013	&	1999-12-30	&	2000-02-05	&	1999-10-09	&	2000-02-26	&	&	 P1012202	&	O5G024NBQ	,	O5G024NCQ	&	v2.2.2	\\
PG1100+772	&	3C249.1	&	0.3114	&	0.034	&	2000-01-20	&	2000-01-31	&1993-05\tablenotemark{f}		&1993-05\tablenotemark{f}		&	&	 P1071601	&	O5G030D4Q	,	O5G030D5Q	&	v2.2.2	\\
PG1259+593	&		&	0.4769	&	0.008	&	2000-02-25	&	2000-02-09	&	2000-02-25	&	2000-02-26	&	&	 P1080101	&	O5G047IJQ	,	O5G047IKQ	&	v2.2.2	\\
PG1322+659	&		&	0.1684	&	0.019	&	2000-05-08	&	2000-06-18	&	2000-02-25	&	2000-02-26	&	&	 A0600808	&	O5G052WXQ	,	O5G052WYQ	&	v2.2.2	\\
PG1351+640	&		&	0.0882	&	0.020	&	2000-01-18	&	1999-10-28	&	2000-02-25	&	2000-02-26	&	&	 P1072501	&	O5G054KQQ	,	O5G054KRQ	&	v2.1.7	\\
PG2349$-$014	&	PKS 2349$-$10	&	0.1740	&	0.027	&	2000-06-25	&	1999-08-27	&	1999-10-07	&	1999-10-08	&	&	 P1074201	&	O5G088N3Q	,	O5G088N4Q	&	v2.2.2	\\
TON951	&	PG0844+349	&	0.0643	&	0.037	&	2000-02-20	&	1999-10-21	&	2000-02-28	&	2000-02-27	&	&	 P1012002	&	O5G020QBQ	,	O5G020QCQ	&	v1.8.7 
\enddata

%\tablecomments{}
\tablenotetext{a}{Measured from the optical data in 
this study (\S\ref{sec:hb}).}
\tablenotetext{b}{From NED based on \citet{Schl98}.}
\tablenotetext{c}{Observed on 1991-10-22, Y0RV0G03T is also used.}
\tablenotetext{d}{X0170101 is also used.}
\tablenotetext{e}{Y0YA0303T and Y0YA0304T are also used.}
\tablenotetext{f}{Obtained with 2.7m telescope at McDonald Observatory.}

\end{deluxetable}

Table~\ref{tb:log} lists the basic parameters and observation log
for the sample.  For more than half of the objects, the \fuse,
\hst, and KPNO spectra were obtained within a few months;  all
but 5 objects (3C351, IRAS~F07546+3928, Mrk509, PG0947+396,
and PG1100+770) were observed in the three bands within a year.
In the case of PG1100+770, we failed to observe the red part of the
spectrum at KPNO in Feb 2000, and have instead used archival (1993)
spectrophotometry obtained with the 2.7 meter telescope at McDonald
Observatory that was consistent with the 2000 epoch blue KPNO
spectrum.  For 3C351 and Mrk509, archival \hst\ data are used, but
the \fuse\ and optical observations were essentially simultaneous,
and can be used to constrain the flux level of the archival \hst\
spectra when necessary.  
Our HST and optical observations of IRAS F07546+3928 were close in
time, but the contemporaneous FUSE observation had now signal due to
wrong pointing of the telescope.
We therefore use a FUSE observation from the FUSE archive obtained
later in time for this object.

%%%%%%%%%%%%%%%%%%%%%%%%%%%%%
\subsection{Optical Spectra}

All the optical spectra were obtained with the 2.1m telescope 
at KPNO except for a few noted in Table~\ref{tb:log}.
A wide slit of 6\arcsec\ was used to ensure that all the 
light from the AGNs is included.  Two spectra were obtained
for each object, covering the observed wavelengths from 
$\sim$3180--6000\AA\ and $\sim$5600--9000\AA, with resolution
of $\sim$9\AA\ and $\sim$12\AA, respectively.

We used standard packages in IRAF to reduce the optical data.
We paid special attention in subtracting the host galaxy contribution
when extracting one-dimensional spectra.  A low-order polynomial was fit
across the dispersion direction within the extracting aperture
to represent the host galaxy contribution and sky background.
The galaxy contribution was clear in many of the lowest luminosity
objects, and appeared to be well subtracted from the final AGN spectra.  
We estimate that the host galaxy contamination is less 
than 5\% in all cases.

Wavelength calibration was done by using comparison spectra, and
absolute flux calibration was achieved by using the standard star spectra
obtained on the same night when it was photometric.  At least one
of the two observations for each object was done under photometric
conditions.  The two spectra were then combined in the observed frame.
Usually, the flux calibration of the two spectra agree very well;
in a few cases when they do not match in the overlap region, we
scale one spectrum to match the one with better flux calibration.

%%%%%%%%%%%%%%%%%%%%%%%%%%%%%%%%%%%%%%%
\subsection{Near Ultraviolet Spectra}

Near-UV (NUV) spectra covering a wavelength range of $\sim$1150--3180\AA\
were obtained from our \hst\ spectroscopic snapshot survey.
Observations were made with STIS in slitless mode, to
minimize target acquisition overheads.  A guide star acquisition
assured good imaging and hence full spectral resolution ($\sim$1\AA) on
these point sources.  
To save more overhead time, we skipped the standard wavelength
calibration observations that are used to determine the zero-point
correction for the wavelength scale. Instead, we used the Galactic
absorption lines clearly visible in each spectrum to determine the
proper zero-point correction. To apply this to the spectra, we
use an iterative process. We run the CALSTIS pipeline with no
zero-point correction, and then measure the wavelengths of the
Galactic
lines. We then calculate the required correction, assuming that the
Galactic lines are at their rest wavelengths, and manually edit the
header information in the CALSTIS pipeline at the point that the
wavelength zero-point correction is made. We then measure the
wavelengths again in the extracted spectra. Usually one iteration is
sufficient, and more than two are never required. The resulting
errors are typically less than 1 \AA.

The spectra taken with G140L and G230L were then combined.
When the two spectra do not match in the wavelength overlap region,
we scaled the G230L spectrum by a uniform factor, which 
typically resulted in a correction less than a few percent.

%%%%%%%%%%%%%%%%%%%%%%%%%%%%%%%%%%%
\subsection{Far Ultraviolet Spectra}

\fuse\ spectra covering observed wavelengths of 905--1187\AA\ have a
high resolution of $\sim$0.05\AA.  Newer versions of the standard
\fuse\ calibration pipeline {\it CalFUSE} were used to process the raw
\fuse/FUV data \citep[see][]{Sahn00}.  Spectra were extracted after
background subtraction with updated background models and subtraction
algorithms.  The wavelength and flux were then calibrated.  The
version of {\it CalFUSE} used for different objects are also listed in
Table~\ref{tb:log} for reference.

Among all the instrumental effects in the \fuse\ data, the most prominent
is the ``worm,'' a dark stripe of decreased flux in the spectra
running in the dispersion direction.  The flux loss can be as much
as 50\% in the longer wavelengths, where we need to connect with
\hst\ spectra.  
To correct for the effects of the ``worm'', we used data from the two
independent LiF channels covering the 1100--1180 \AA\ wavelength
range. The worm is most often present in channel LiF1b. After
confirming this by inspection, we form a ratio of the spectra
obtained in the two independent channels. Assuming that the data
in channel LiF2a are uncorrupted, we fit a low-order spline to the
ratio to derive a correction curve for LiF1b, and apply this to the
LiF1b data. This removes the ``worm'' while preserving the
high-spectral resolution and statistical independence of
the LiF1b data.  More information can
be found at the \fuse\
webpage\footnote{\url{http://fuse.pha.jhu.edu/analysis/calfuse.html}}.

\subsection{Soft X-ray data}

We do not have simultaneous X-ray observations.  Instead, we collected
available soft X-ray spectral indices and fluxes (0.1--2.4 \kev,
Table~\ref{tb:index}) from the literature \citep{Brin97,Pfef01}.
These data were obtained from both {\it ROSAT} All-Sky Survey (RASS) and
public PSPC observations.  Assuming Galactic absorption, the power-law X-ray
photon indices were estimated from the two hardness ratios given
by the Standard Analysis Software System \citep{Brin97} or from a spectral fit \citep{Pfef01}.
X-ray fluxes between 0.1--2.4\,keV were also obtained.  We used the
photon index and the X-ray flux to calculate the flux density
at 1\,keV and the X-ray spectral index \ax, and used them in the
correlation analysis (\S\ref{sec:corr}).  

We did not find X-ray information for PG~1259+593, and the X-ray flux
of Mrk~304 is too low to derive a spectral index \citep{Brin97}.
There are Chandra observations of NGC~3783 \citep[e.g.,][]{Kasp02},
but the data show strong variability and warm absorbers
that complicate a consistent comparison with the spectral indices 
derived from the ROSAT observations for other objects.

\section{CONSTRUCTION OF SPECTRAL ENERGY DISTRIBUTIONS\label{sec:construction}}

The very strong geocoronal emission lines, i.e. airglow emission
(e.g., \oil, \lyal, \lybl\ etc.), were first removed by hand from both
the \fuse\ and \hst\ spectra.  We made no attempt to remove numerous
narrow Galactic molecular hydrogen absorption lines.  The \fuse, \hst, and
optical spectra were then combined.  The overlap regions between the
spectra are usually 20--100\AA, large enough for us to assess the
agreement of flux level in the spectra due to flux calibration, source
variability, and ``worm'' correction for \fuse\ spectra.  The overlap
region between the \hst\ and optical spectra is relatively small, and
sometimes there is a gap of 5--40\AA.  However, we found that this
does not prevent us from matching the continuum levels on both sides
of the gap, because a flux change of 5\% at the gap would be very
obvious by checking the overall continuum trend on both sides of the
gap.

The flux densities of the  optical and \hst\ spectra usually agree
very well within 3\%, and we do not scale the
spectra to match in this case because the uncertainty in the scaling
factor is at a similar level.  When the flux density of the \fuse\ 
spectrum does not agree with the above, we scale it to match 
using a constant scaling factor determined from the overlap regions
between \fuse\ and \hst\ spectra (Table~\ref{tb:gap}).  In the cases when
the \fuse, \hst, and optical spectra do not agree with each other, we
scale the \fuse\ and optical spectra to match the \hst\ spectrum,
because based on our knowledge, the flux calibration of the \hst\ STIS
spectra is usually more reliable.  Most flux differences are slight
($<20\%$), and corrections in the 20--40\% range affect less than a
quarter of the sample.  Since the spectral shape, rather than the
absolute flux, is the most important in constructing SEDs, this
scaling process should not significantly affect our study of SEDs if
problems in the flux calibration are the cause (e.g., because of
clouds).

However, based on our knowledge of the flux calibrations, we
believe the change of the continuum levels in different wavebands is
most likely due to source variability.  Although we tried to obtain
quasi-simultaneous spectra for each object, some observations were
actually separated by a few months, and the continuum level could vary
by a factor of 20\% or more, and it is empirically true that most AGNs
show spectral shape changes with flux changes.  We list the observing
time gaps and scaling factors applied for the spectra in
Table~\ref{tb:gap} as a guide to assess this problem.

\begin{deluxetable}{lrrrrrrrrr}
\tabletypesize{\scriptsize}
%\tabletypesize{\tiny}
%\rotate
\tablecaption{Observing Time Gaps and Scaling Factors \label{tb:gap}}
\tablewidth{0pt} 
\tablehead{ 
\colhead{} &
\multicolumn{4}{c}{Observing Time Gap (days)\tablenotemark{a}} &
\colhead{} &
\multicolumn{4}{c}{Scaling Factor} 
\\ \cline{2-5} \cline{7-10}
\colhead{Object} &
\colhead{\fuse} &
\colhead{\hst} &
\colhead{opt. blue} &
\colhead{opt. red} & &
\colhead{\fuse} &
\colhead{\hst} &
\colhead{opt. blue\tablenotemark{b}} &
\colhead{opt. red\tablenotemark{b}} 
}

\startdata

3C273	&	38	&	0	&	$-$20	&	$-$19
&	&	1.03	&	1	&	1	&	0.94
\\
3C351	&	0	&	$-$2801	&	$-$8	& $-$2288
&	&	1.10	&	1	&	0.76	&	scaled
\\
% 3C351	&	2801	&	0	&	2793	&	$-$513
% &	&	1.10	&	1	&	0.76	&	scaled
% \\
4C+34.47	&	$-$16	&	0	&	$-$260	&
$-$119	&	&	1.13	&	1	&	1	&
1	\\
IR07546+3928	&	744	&	0	&	$-$112	&
$-$113	&	&	1.12	&	1	&	1	&
0.95	\\
MRK290	&	$-$78	&	0	&	$-$98	&	$-$96
&	&	1.33	&	1	&	1	&	0.73
\\
MRK304	&	27	&	0	&	$-$256	&	$-$255
&	&	0.99\tablenotemark{c}	&	1	&	1	&	0.95
\\
MRK506	&	$-$16	&	0	&	$-$259	&	$-$118
&	&	1.09	&	1	&	1\tablenotemark{d}
&	1\tablenotemark{d}
\\
MRK509	&	0	&	$-$2694	&	35	&	35
&	&	1.01\tablenotemark{c}	&	1	&	1
&	1.03\tablenotemark{c}
\\
% MRK509	&	2694	&	0	&	2729	&	2729
% &	&	1.01\tablenotemark{c}	&	1	&	1
% &	1.03\tablenotemark{c}
% \\
PG0052+251	&	2	&	0	&	6	&
7	&	&	1.01\tablenotemark{c}	&	1	&	1	&
1	\\
PG0947+396	&	205	&	0	&	$-$250	&
$-$109	&	&	0.85	&	1	&	1	&
scaled	\\
PG0953+414	&	$-$37	&	0	&	$-$119	&
21	&	&	0.67	&	1	&	1	&
scaled	\\
PG1100+772	&	$-$11	&	0	&	$\sim$$-$2450	&
$\sim$$-$2450	&	&	0.91	&	1	&	1.10	&
1.10	\\
PG1259+593	&	16	&	0	&	16	&
17	&	&	0.77	&	1	&	1	&
1	\\
PG1322+659	&	$-$41	&	0	&	$-$114	&
$-$113	&	&	1.04	&	1	&	0.79	&
1	\\
PG1351+640	&	82	&	0	&	120	&
121	&	&	1.00	&	1	&	1.67	&
1.50	\\
PG2349$-$014	&	303	&	0	&	41	&
42	&	&	0.44	&	1	&	1	&
1	\\
TON951	&	122	&	0	&	130	&	129
&	&	0.82	&	1	&	1	&	1

\enddata

\tablenotetext{a}{Relative to \hst\ observing time except for
3C351 and Mrk509, for which \hst\ archival data are used and the
time gap is relative to \fuse\ observing time.}

\tablenotetext{b}{When the optical blue spectra match the \hst\
spectra within a few percent, the same level as the uncertainty, 
we do not apply a scaling (scaling
factor=1).  Optical red spectra are often not photometric.
Scaling the red to match the blue is not a problem since they were
usually obtained within 1--2 days.}

\tablenotetext{c}{No scaling is applied when combining the spectra
since the scaling factor is small.}

\tablenotetext{d}{Data were photometric, but host galaxy contribution
is large.  We removed the galaxy contribution until everything matches
properly.  The correction of galaxy is good to 10\%.
}

\end{deluxetable}

We corrected for Galactic reddening with an empirical mean
extinction law \citep*[][CCM]{Card89}, assuming $R_V = A_V/E(\bv)
= 3.1$, a typical value for the diffuse interstellar medium.  $E(\bv)$
is obtained from NED\footnote{NASA/IPAC Extragalactic Database (NED)
is operated by the Jet Propulsion Laboratory, California Institute of
Technology, under contract with the National Aeronautics and Space
Administration, \url{http://nedwww.ipac.caltech.edu/}} based on the
dust map created by \citet{Schl98}.  Since the \citet{Card89} extinction
curve's UV cutoff is at 1000\AA, we extrapolate the curve
down to 900\AA\ for our FUV spectra by using the same formula.
This extrapolation of CCM law to shorter wavelength has been shown to
be compatible with recent FUSE data \citep[e.g.,][]{HutGia01}.

We finally applied the redshift correction and brought both the
wavelength (in vacuum) and flux density (\flambda) of the spectra to
rest frame.  The redshifts were determined using measurements of
\oiiil\ in our optical data (\S\ref{sec:hb}).

The final SEDs of this sample are shown in Figure~{\ref{fg:sed}}.  All
the spectra of the 19 AGNs (including NGC~3516 and NGC~3783, but not
NGC~5548) are available at {\verb+http://physics.uwyo.edu/agn/+}.
These include the unscaled individual \fuse, \hst, and optical spectra
as well as the combined SEDs.  The special treatment of wavelength
calibration in the \hst\ spectra is not a problem for using the data for
other studies, e.g., emission-line analyses,
---
the wavelength calibrations are only slightly less accurate than
standard STIS observations (0.5--1.0 pixels rms vs.  0.1--0.2 pixels),
and this has no measurable effect on the flux calibration.

%\clearpage    
\begin{figure}
\epsscale{.80}
\plotone{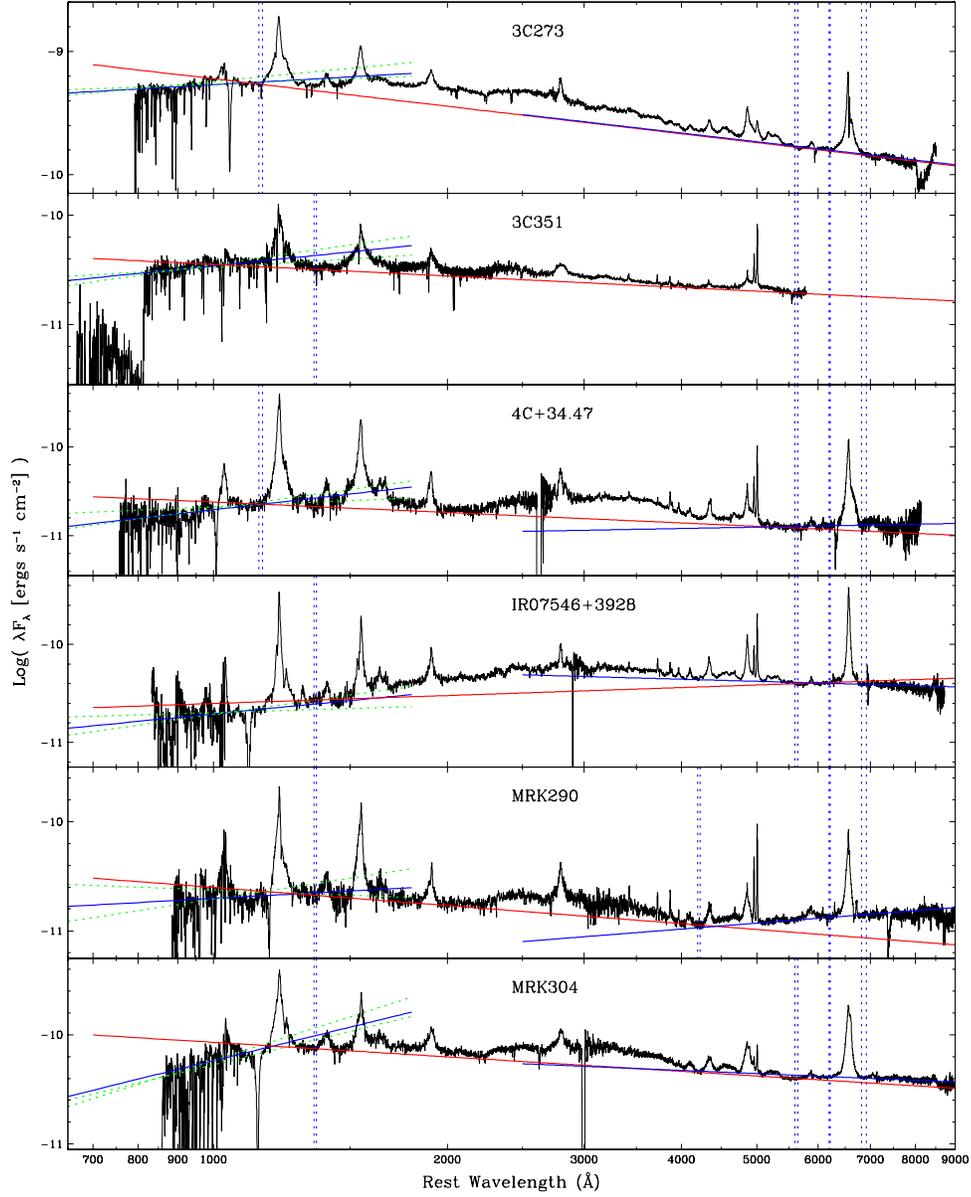}
\caption{FUV to optical SEDs and fitted power-laws for
different regions.  Uncertainties in the FUV power-law
fitting are also shown.  The vertical dotted lines indicate
the continuum windows used for fitting 
\auvoa\ ($\sim$1200--5500\AA) and 
\auvob\ ($\sim$5500--9000\AA) (Table~\ref{tb:index}).  
FUSE spectra have been rebinned to a resolution of 0.5\AA\
for display purpose.
\label{fg:sed}}
\end{figure}   

%\clearpage

\addtocounter{figure}{-1}
%\clearpage    
\begin{figure}
\plotone{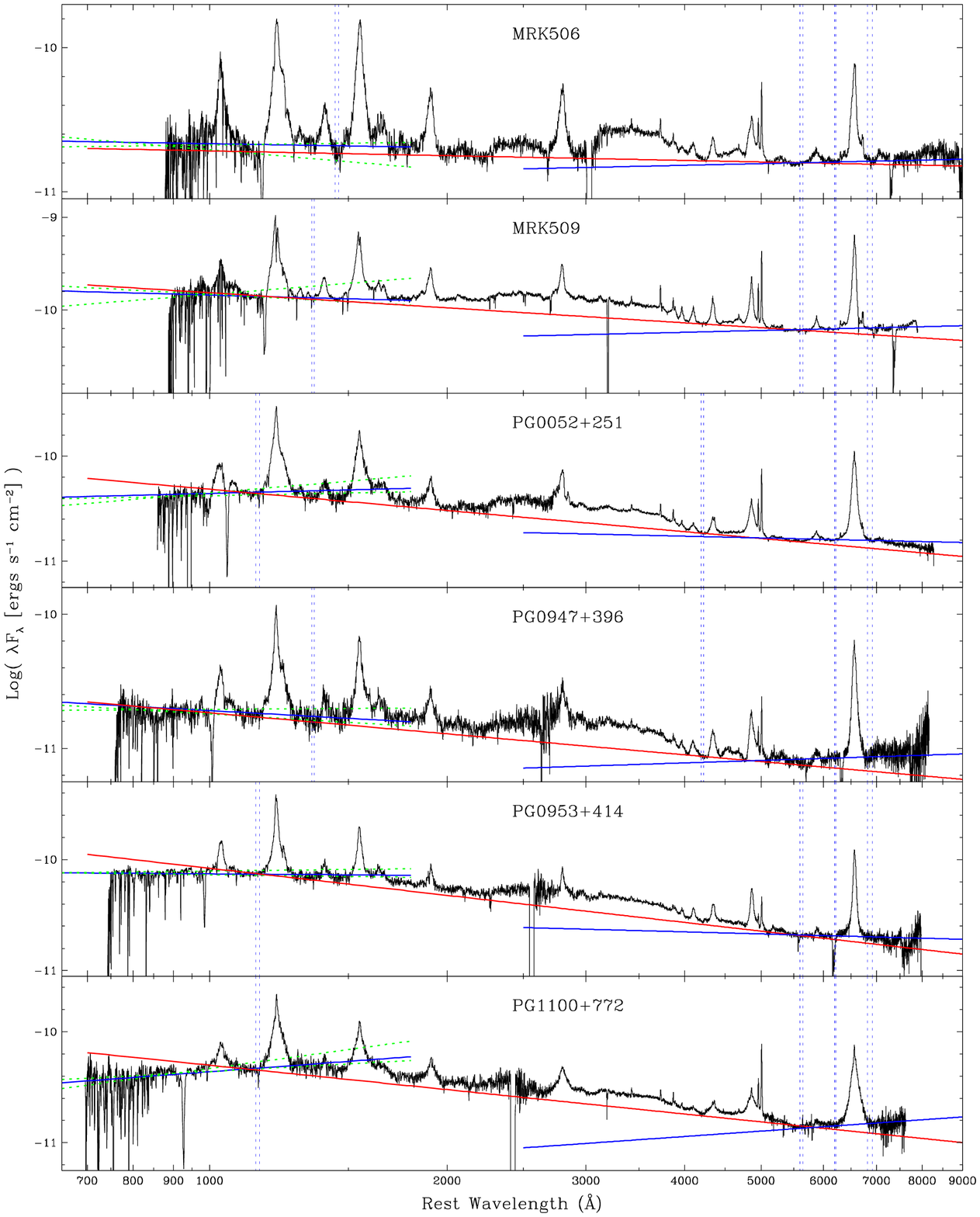}
\caption{Continued
\label{fg:sed2}}
\end{figure}   %\clearpage

\addtocounter{figure}{-1}
%\clearpage    
\begin{figure}
\plotone{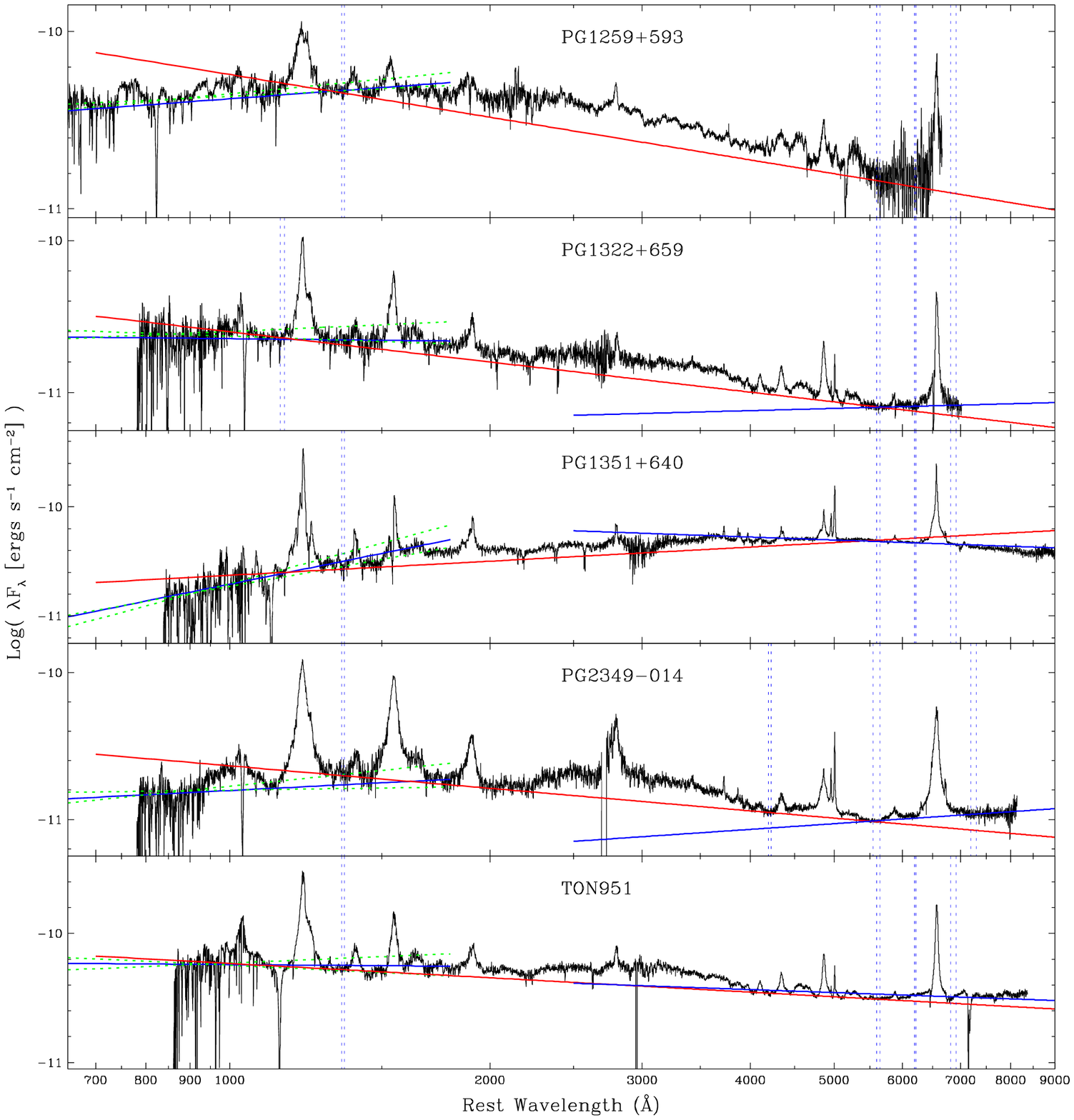}
\caption{Continued
\label{fg:sed3}}
\end{figure}   %\clearpage

%\clearpage

\section{SED ANALYSIS}

\subsection{SED Measurements\label{sec:cont}}

For many objects, the FUV and NUV-optical spectral
regions have different slopes (Fig.~\ref{fg:sed}).  After careful
examination, we decided to use three power-laws to fit the entire
FUV-optical continuum.

A power-law with spectral index of \afuv\ is fitted to the FUV
spectral region ( $\lesssim$1100\AA)  on a case-by-case basis.  We
first exclude the obvious emission-line regions (\lygl, {C\,{\sc
iii}\,$\lambda$977, {N\,{\sc iii}\,$\lambda$991, \lybl, \ovil,
\siiv\,$\lambda$1062, \siiv\,$\lambda$1073, and \heii\,$\lambda$1084,
etc.)  Strong ISM Lyman series absorption lines and possible AGN \ovi\ and
\lya\ absorption features are also excluded.  We make sure the
excluded regions are wide enough by visual inspection so that the line
wings are also excluded.  We then fit the remaining data points with a
power-law, and iteratively reject points beyond $\pm 2\sigma$ and the
points next to the rejected points.  This process removes
the numerous narrow Galactic H$_2$ absorption lines and possible
residual emission features from the fitting.  Finally, we visually
inspect the fitting results and make sure that the power-law goes
through the apparent continuum regions in the FUV spectra.  We repeat
this process to estimate upper and lower limits for \afuv\ by
intentionally including some emission or absorption features until the
fitted power-law obviously deviates from the spectra (visual inspection).
We therefore obtain conservative uncertainties 
for the best-fit power-law.
The fitted \afuv\ and its uncertainties are listed in
Table~\ref{tb:index}.  If there are possible residual weak blended
Galactic absorption lines or emission features, they are only
comparable with the noise level, and the uncertainty in the fitted
spectra indices caused by these weak features is much
smaller than the above estimated uncertainties.

\begin{deluxetable}{lrrcrcrcrrlrr}
%\tabletypesize{\scriptsize}
\tabletypesize{\tiny}
%\rotate
\tablecaption{Spectral Indexes\label{tb:index}}
\tablewidth{0pt}
\tablehead{ 
%\colhead{} & \colhead{} & \colhead{} & \colhead{} &  
%\multicolumn{5}{c}{Observation Date}  &
%\multicolumn{2}{c}{Dateset ID} &
%\colhead{CalFUSE} \\ \cline{5-8} \cline{10-11}
\colhead{} &
\multicolumn{2}{c}{FUV} &
\colhead{Cont. Windows\tablenotemark{b}} &
\multicolumn{2}{c}{1200--5500\AA} &
\colhead{} &
\multicolumn{2}{c}{5500--9000\AA} &
\colhead{} & \colhead{} &
\multicolumn{2}{c}{0.1--2.4 keV}
\\ \cline{2-3} \cline{5-6} \cline{8-9} \cline{12-13}
\colhead{Object} & 
\colhead{$f_{1000}$\tablenotemark{a}} & 
\colhead{\afuv} & 
\colhead{a b c d e f} & 
\colhead{$f_{1000}$\tablenotemark{a}} & 
\colhead{\auvoa} & 
\colhead{} &
\colhead{$f_{1000}$\tablenotemark{a}} & 
\colhead{\auvob} & 
\colhead{$\Delta\alpha$\tablenotemark{c}} & 
\colhead{$\lambda_{break}$\tablenotemark{d}} & 
\colhead{$f_{\rm 1kev}$\tablenotemark{e}} & 
\colhead{\ax\tablenotemark{e}}  
}
\startdata
% obj	&	fuvf1000	&	     fuva
% &	a		b		c		d
% e		f	&	uvo1f1000	&	  uvoa1	&
% &	uvo2f1000	&	   uvoa2	&	diffa	&
% wavebreak	wavebreakerr1	wavebreakerr2	&	xf1kev	&
% ax	 axe1	  axe2	\\
3C273	&	53.70	&	$-0.64^{+0.24}_{-0.11}$	&	x
\phm{x}		\phm{x}		x		x		x
&	60.00	&	$-$1.74	&	&	60.00	&	$-$1.73
&	1.10	&	$1105^{+0}_{-69}$	&	283.0	&
$-1.11^{+0.01}_{-0.01}$	\\
3C351	&	3.44	&	$-0.28^{+0.30}_{-0.28}$	&
\phm{x}		x		\phm{x}		x		x
x	&	4.69	&	$-$1.35	&	&	\nodata	&
\nodata	&	1.07	&	$1027^{+57}_{-25}$	&	3.1
&	$-1.31^{+0.05}_{-0.05}$	\\
4C+34.47	&	1.95	&	$0.00^{+0.21}_{-0.61}$	&
x		\phm{x}		\phm{x}		x		x
x	&	2.39	&	$-$1.39	&	&	0.97	&
$-$0.84	&	1.39	&	$1155^{+16}_{-45}$	&	66.3
&	$-1.29^{+0.06}_{-0.06}$	\\
IRAS F07546...	&	1.95	&	$-0.22^{+0.35}_{-0.54}$	&
\phm{x}		x		\phm{x}		x		x
x	&	2.50	&	$-$0.73	&	&	5.99	&
$-$1.22	&	0.51	&	$1620^{+1619}_{-266}$	&	6.9
&	$-2.16^{+0.38}_{-0.38}$	\\
MRK290	&	1.99	&	$-0.61^{+0.70}_{-0.66}$	&
\phm{x}		x		x		\phm{x}		x
x	&	2.50	&	$-$1.55	&	&	0.48	&
$-$0.44	&	0.94	&	$1281^{+69}_{-118}$	&	30.1
&	$-1.32^{+0.13}_{-0.13}$	\\
MRK304	&	5.78	&	$0.75^{+0.52}_{-0.01}$	&
\phm{x}		x		\phm{x}		x		x
x	&	8.52	&	$-$1.44	&	&	7.15	&
$-$1.30	&	2.18	&	$1193^{+49}_{-43}$	&
\nodata
&	   \nodata			\\
MRK506	&	2.15	&	$-1.09^{+0.16}_{-0.38}$	&	-
\phm{x}		\phm{x}		x		x		x
&	1.92	&	$-$1.11	&	&	1.29	&	$-$0.88
&	0.02	&	\nodata			&	26.6	&
$-1.20^{+0.09}_{-0.09}$	\\
MRK509	&	14.50	&	$-1.23^{+0.91}_{-0.23}$	&
\phm{x}		x		\phm{x}		x		x
x	&	15.30	&	$-$1.54	&	&	4.32	&
$-$0.80	&	0.31	&	$1201^{+358}_{-167}$	&	386.0
&	$-1.61^{+0.03}_{-0.03}$	\\
PG0052+251	&	4.41	&	$-0.81^{+0.45}_{-0.04}$	&
x		\phm{x}		x		\phm{x}		x
x	&	4.84	&	$-$1.67	&	&	2.19	&
$-$1.17	&	0.86	&	$1111^{+78}_{-48}$	&	46.2
&	$-1.49^{+0.02}_{-0.04}$	\\
PG0947+396	&	1.92	&	$-1.33^{+0.37}_{-0.02}$	&
\phm{x}		x		x		\phm{x}		x
x	&	1.84	&	$-$1.52	&	&	0.60	&
$-$0.81	&	0.19	&	$\phn813^{+323}_{-87}$	&	14.9
&	$-1.18^{+0.18}_{-0.15}$	\\
PG0953+414	&	7.43	&	$-1.05^{+0.13}_{-0.04}$	&
x		\phm{x}		\phm{x}		x		x
x	&	8.35	&	$-$1.81	&	&	2.91	&
$-$1.19	&	0.76	&	$1165^{+39}_{-100}$	&	17.0
&	$-1.43^{+0.03}_{-0.05}$	\\
PG1100+772	&	4.38	&	$-0.47^{+0.44}_{-0.14}$	&
x		\phm{x}		\phm{x}		x		x
x	&	4.99	&	$-$1.73	&	&	0.57	&
$-$0.50	&	1.26	&	$1114^{+10}_{-72}$	&	12.5
&	$-1.56^{+0.08}_{-0.10}$	\\
PG1259+593	&	4.17	&	$-0.64^{+0.11}_{-0.12}$	&
\phm{x}		x		\phm{x}		x		x
x	&	5.72	&	$-$1.80	&	&	\nodata	&
\nodata	&	1.16	&	$1313^{+1}_{-93}$	&
\nodata
&	   \nodata			\\
PG1322+659	&	2.26	&	$-1.06^{+0.31}_{-0.14}$	&
x		\phm{x}		\phm{x}		x		x
x	&	2.51	&	$-$1.66	&	&	0.62	&
$-$0.85	&	0.60	&	$1191^{+36}_{-196}$	&	17.8
&	$-1.75^{+0.04}_{-0.03}$	\\
PG1351+640	&	1.96	&	$0.60^{+0.51}_{-0.20}$	&
\phm{x}		x		\phm{x}		x		x
x	&	2.35	&	$-$0.57	&	&	7.80	&
$-$1.28	&	1.17	&	$1173^{+101}_{-62}$	&	4.2
&	$-1.43^{+0.06}_{-0.06}$	\\
PG2349-014	&	1.57	&	$-0.70^{+0.31}_{-0.22}$	&
\phm{x}		x		x		\phm{x}		-
-	&	2.32	&	$-$1.51	&	&	0.49	&
	$-$0.60	&	0.81	&	$1616^{+292}_{-284}$	&
31.2	&	$-1.44^{+0.12}_{-0.12}$	\\
TON951	&	5.74	&	$-1.05^{+0.33}_{-0.27}$	&
\phm{x}		x		\phm{x}		x		x
x	&	5.86	&	$-$1.37	&	&	5.12	&
$-$1.24	&	0.32	&	$1066^{+880}_{-77}$	&	3.2
&	$-1.54^{+0.11}_{-0.12}$	
\enddata

%\tablecomments{}
\tablenotetext{a}{$f_{1000}$ --- fitted rest frame continuum flux density
at 1000\AA\ for corresponding regions ($10^{-14}$ \fluxl).}
\tablenotetext{b}{Continuum windows used for fitting NUV-optical region.
a:1144--1157;
b:1348--1358;
c:4200--4230;
d:5600--5648;
e:6198--6215;
f:6820--6920.
For Mrk506, 1444--1458\AA\ is used instead of a; for PG2349$-$014, 
5550--5650\AA\ and 7200--7300\AA\ are used instead of e and f.
}
\tablenotetext{c}{$\Delta\alpha = \afuv - \auvoa$}
\tablenotetext{d}{Break wavelength for \afuv\ and \auvoa.  The errors
are calculated solely from the errors of \afuv\ since the uncertainty
of \auvoa\ is negligible based on the way it is measured (\S\ref{sec:cont}).}
\tablenotetext{e}{$f_{\rm 1kev}$ --- $f_\lambda$ at 1\,keV ($10^{-14}$ \fluxl).  
Calculated from integrated flux between 0.1--2.4\,keV and \ax\ 
($\fnu \propto \nu^{\ax}$) from
\citet{Brin97} and \citet[][for Mrk290 and Mrk506 only]{Pfef01}, 
assuming power-law.}

\end{deluxetable}

A single power-law cannot fit the entire NUV-optical region
in many objects.  This was noticed before, for example, in the Sloan Digital Sky
Survey (SDSS) composite
spectra \citep{Vand01}.  Clean continuum regions are also hard
to find in the AGN NUV-optical spectra due to the large number of
broad emission lines and blends, including the ``small blue bump''
from $\sim$2000--4000\AA--- the blend of \feii\ emission and Balmer
continuum.  We therefore have to define, for this sample, some
common, narrow continuum windows where there seem to be no emission lines:
1144--1157\AA,
1348--1358\AA,
4200--4230\AA,
5600--5648\AA,
6198--6215\AA, and 
6820--6920\AA.

The NUV-optical spectra index \auvoa\ in $\sim$1200--5500\AA, and red
optical spectra index \auvob\ in $\sim$5500--9000\AA\ are each
obtained by fitting a power-law to a pair of selected continuum
windows, requiring that all emission features are above the fitted
power-laws in the corresponding regions.  The continuum windows used
for each object are listed in Table~\ref{tb:index} and marked in
Figure~\ref{fg:sed}.  Since the continuum windows are very narrow,
this fitting process is more like defining each power-law with two
points.  The power laws cannot be treated as the true continua of the
spectra, but the spectral indices provide information on the overall
continuum slopes.

The UV bump can now be characterized with two power-laws (a broken power-law)
with spectral indices of \afuv\ and \auvoa, and a break wavelength
\lbreak,
which is defined by the intersection of the two power-laws.

%%%%%%%%%%%%%%%%%%%%%%%%%%%%%%%%%%%%%%%%%%%%%%%%%%%%
\subsection{Spectral Break in SEDs\label{sec:break}}

As can be seen in Figure 1, an extrapolation of the NUV-optical
power-law does not match the FUV continuum in most objects. For the
possible exceptions, Mrk 290, Mrk 506, Mrk 509, PG0947+396, and Ton
951, the extrapolated NUV-optical power-law falls within the bounds of
the errors for our fits to the FUV continuum. Thus, for 12 out of 17
objects, we see a break in the spectral index to a steeper value when
comparing the NUV-optical to the FUV continuum.

% As can be seen in Figure~\ref{fg:sed}, a power-law continuum for
% the NUV-optical region cannot be extrapolated to match the FUV continuum
% in most objects.  The possible exceptions are Mrk506, Mrk509,
% PG0947+396, and Ton951, in which the FUV region seems to be above the
% NUV-optical power-law fitting, but detailed fitting still shows that
% their FUV spectra have different slopes from those of NUV-optical
% region except for Mrk506.  A spectral break is clearly seen for most
% objects.  In the case of Mrk506, its \afuv\ and \auvoa\ are about the
% same, so no spectral break is seen.

The break wavelength is calculated as the intersection point of
the two power-laws.  The
distribution of the break wavelength is shown in
Figure~\ref{fg:wavehist}, where \lbreak\ peaks near 1100\AA\ and spans
800--1600\AA.  However, since the calculated break wavelength
is very sensitive to small changes in \afuv\ or \auvoa\ when the
difference between \afuv\ and \auvoa\ is small,  it has a large
uncertainty for some objects (Table~\ref{tb:index}).

% The largest \lbreak\ ($>$1400\AA) are for
% \irasobj\ and PG2349-014.  \irasobj\ has evidence of being
% intrinsically reddened (\S\ref{sec:disc}).

%\clearpage    
\begin{figure}[t]
%\epsscale{.80}
\includegraphics[angle=270,scale=0.3]{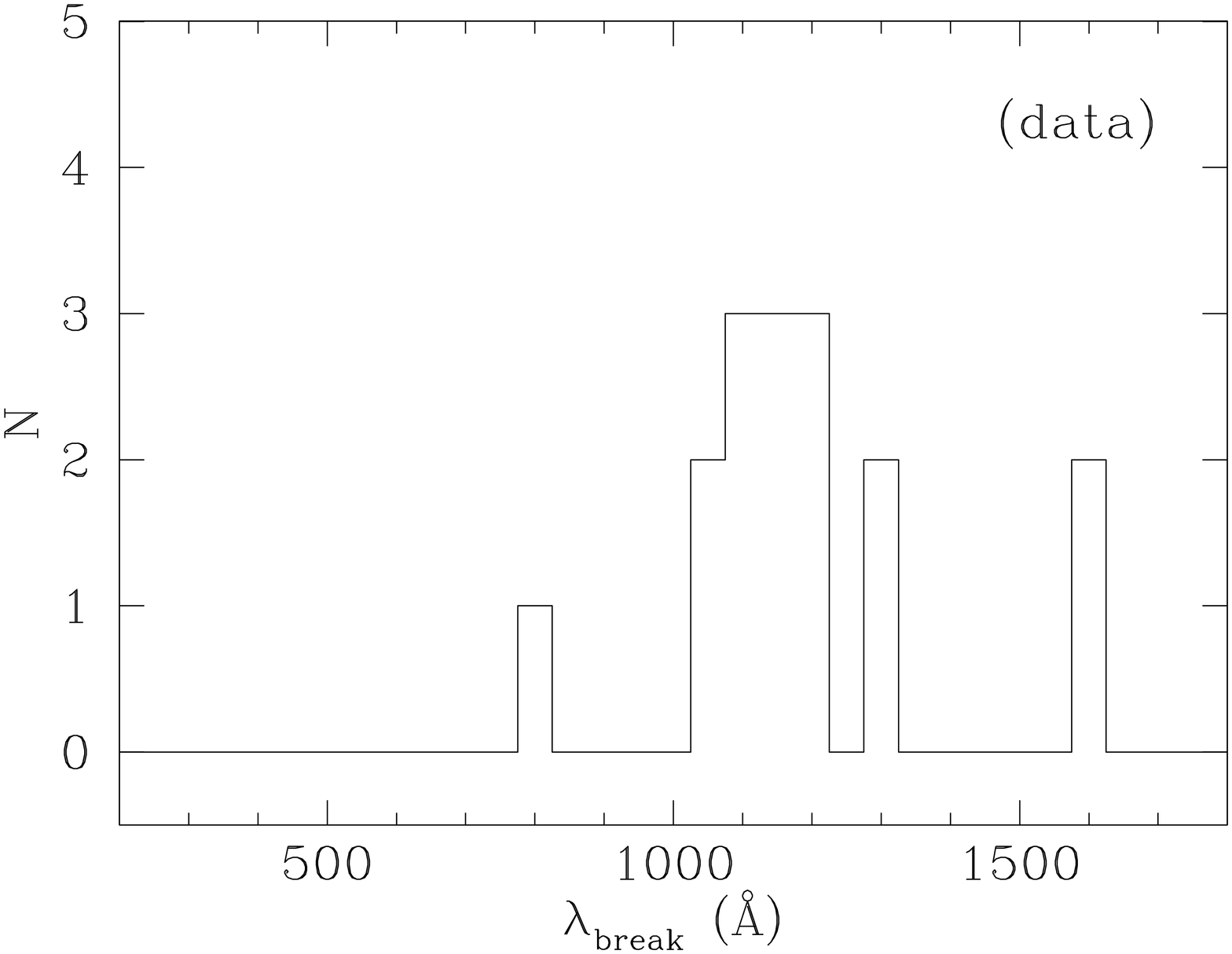}
\caption{Distribution of the break wavelength, indicating
a spectral break near 1100\AA\ for most objects.  The 
largest \lbreak\ is from \irasobj\ (\S\ref{sec:break}) and PG2349-014.
Mrk506 does not show a spectral break and is not included.
\label{fg:wavehist}}
\end{figure}   %\clearpage

We have also compared our SEDs with the soft X-ray spectral indices
in Figure~\ref{fg:xsed} in a similar way as in \citet{Laor97},
except that we also have FUV data.  For more than half our objects,
the soft X-ray spectral indices appear to match up reasonably with
the extrapolation of the FUV continuum.  This directly confirms the
finding by \citet{Laor97} and \citet{Zhen97} in composite spectra
that the peak of the big blue bump lies
in the FUV region. 

Three objects, IRAS F07546+3928, NGC~3516, and PG1351+640, show
strongly suppressed NUV-FUV continua. These objects also show
intrinsic absorption features, possibly suggesting the existence
of dust associated with the absorbers.
(See \citet{Zhen01} for the case of PG~1351+640.)
We will discuss the reddening effect more in \S\ref{sec:disc}.

% Here we can also see that the NUV-FUV spectra of \irasobj\ has been
% strongly suppressed, an indication of possible large internal
% reddening.  Two other objects with obviously suppressed UV continuum
% are NGC~3516 and PG~1351+640, both of which also show intrinsic
% absorption features, possibly suggesting the existence of dust
% associated with the absorbers.  

%\clearpage    
\begin{figure}
\epsscale{.80}
\plotone{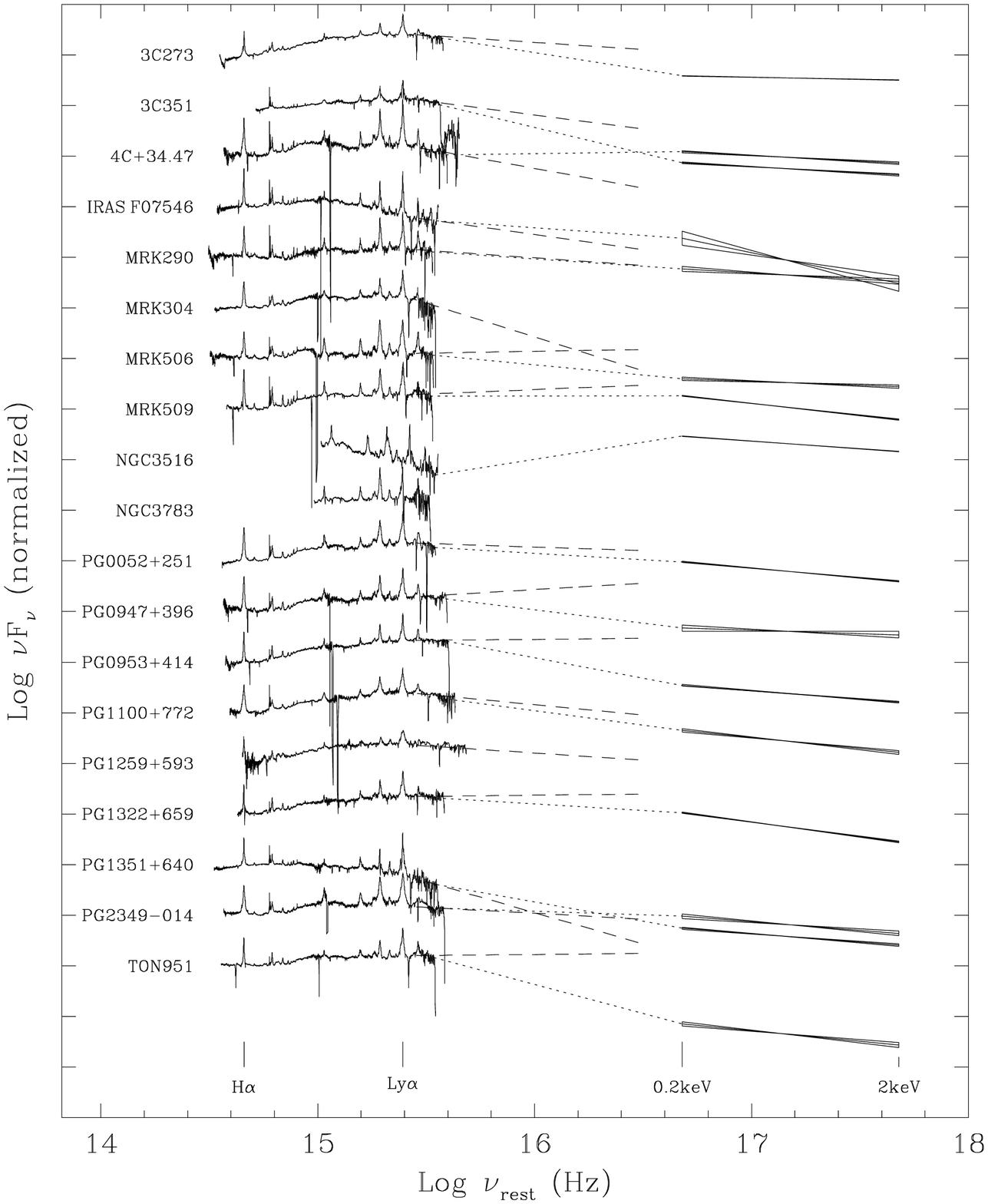}
\caption{FUV-optical SEDs and the soft X-ray spectral indices.
The dotted lines are drawn to connect FUV spectra and X-ray spectral
slopes, and the dashed lines indicate the extrapolation of fitted FUV
power-law.
More than half of the objects have an X-ray spectral slope roughly matching
the extrapolation of the FUV continuum slope.  Also note the
large UV suppression of \irasobj, NGC~3516, and PG1351+640.
\label{fg:xsed}}
\end{figure}   %\clearpage

%%%%%%%%%%%%%%%%%%%%%%%%%%%%%%%%%%%%%%%%%%%%%%%%%%%%%%
\subsection{Bolometric Luminosity  \label{sec:lbol}}

Bolometric luminosity (\lbol) is one of the most fundamental parameters
for understanding the black hole accretion in quasars, however, it has
not been easy to obtain for quasars in general, because they emit
significant power over a large part of the electromagnetic spectrum.

\citet{Elvi94} built SEDs for a sample of 47 quasars and were able to
obtain the bolometric luminosity by integrating over the SEDs from
radio to X-ray wavelengths.  They also determined the bolometric
correction factors for a few monochromatic luminosities, e.g., $\lbol
\approx 13 \lllambda$(5400\AA), based on their SEDs.
Different empirical correction
factors have been determined and used to estimate \lbol\ in previous
studies \citep[e.g.,][]{Sand89,LaoDra93,Wand99}.  Recently, many studies
use the prescription of
\citet{Kasp00} to estimate quasar black hole masses and bolometric
luminosity $\lbol=9\lllambda$(5100\AA).

Since we have broad spectral coverage from the FUV to optical, we are able
to obtain an accurate luminosity for this region by integrating over
the power-laws we measure.  To estimate the bolometric luminosity, we
need to include X-ray and IR regions.  We extend our FUV
power-law continuum to 700\AA, and then use a power-law to connect
with the soft X-ray luminosity at 0.2\,keV.  

The case for the infrared region is more complicated.  There is an IR
bump around 10\micron\ in quasar SEDs, which also contributes a
significant amount of energy.  After averaging the mean SEDs for
radio-loud and radio-quiet objects from \citet{Elvi94}, we fit two
power-laws to characterize this bump and obtain $\alphal=-0.69$ for
1--10\micron\ and $\alphal=-1.65$ for 10--100\micron.  These two
power-laws form a peak at 11.45\micron, roughly corresponding to the
peak in the mean SEDs.  We scale this power-law IR bump to match the
extrapolation of the fitted NUV-optical 
continuum at 1\micron\ for each object. 

We estimate the bolometric luminosity between 2\,keV and 100\micron\ by
integrating this set of power-laws over this region.  We have also
obtained the luminosities for individual wavebands.
Table~\ref{tb:lbol} lists the results.

\begin{deluxetable}{lrrrrcrrcrrrr}
\tabletypesize{\scriptsize}
%\tabletypesize{\tiny}
%\rotate
\tablecaption{Integral Bolometric and Individual Waveband Luminosities\label{tb:lbol}}
\tablewidth{0pt}
\tablehead{
%\colhead{} & \colhead{} & \colhead{} & \colhead{} &  
%\multicolumn{5}{c}{Observation Date}  &
%\multicolumn{2}{c}{Dateset ID} &
%\colhead{CalFUSE} \\ \cline{5-8} \cline{10-11}
\colhead{Object} &
\colhead{\Lx} &
\colhead{\Lxfuv} &
%\colhead{\Lfuv} &
\multicolumn{2}{c}{\Lfuvo} &
\colhead{} &
\multicolumn{2}{c}{\Lir} &
\colhead{} &
\multicolumn{4}{c}{\Lbol} 
\\ \cline{4-5} \cline{7-8} \cline{10-13}

\colhead{} &
\colhead{} &
\colhead{} &
\colhead{case A\tablenotemark{a}} &
\colhead{case B\tablenotemark{a}} &
\colhead{} &
\colhead{case A\tablenotemark{a}} &
\colhead{case B\tablenotemark{a}} &
\colhead{} &
\colhead{case A\tablenotemark{a}} &
\colhead{R$_A$\tablenotemark{b}} &
\colhead{case B\tablenotemark{a}} &
\colhead{R$_B$\tablenotemark{b}} 
}
\startdata

3C273        &45.51&46.22&46.51&46.51&&46.42&46.41&&46.94& 1.48&46.94&
1.47 \\ 
3C351        &44.23&45.44&   \nodata&46.04&&   \nodata&46.20&&   \nodata&   \nodata&46.50&
1.03 \\ 
4C+34.47     &45.13&45.30&45.42&45.41&&45.72&45.56&&46.07& 1.62&46.00&
1.39 \\ 
IR07546+3928 &43.83&44.54&45.09&45.11&&45.53&45.64&&45.72& 0.98&45.80&
1.17 \\ 
MRK290       &43.23&43.64&43.86&43.81&&44.26&43.86&&44.52& 1.83&44.34&
1.20 \\ 
MRK304\tablenotemark{c}       &   \nodata&   \nodata&45.02&45.01&&45.22&45.16&&45.46& 1.08&45.42&
0.99 \\ 
MRK506       &43.44&43.95&44.20&44.19&&44.54&44.48&&44.81& 1.39&44.78&
1.29 \\ 
MRK509       &44.53&44.87&44.75&44.74&&44.96&44.77&&45.44& 2.28&45.38&
1.99 \\ 
PG0052+251   &44.81&45.33&45.43&45.42&&45.53&45.37&&45.98& 1.83&45.93&
1.62 \\ 
PG0947+396   &44.45&45.11&45.31&45.29&&45.54&45.32&&45.86& 1.71&45.76&
1.36 \\ 
PG0953+414   &44.67&45.66&45.94&45.94&&45.94&45.78&&46.41& 1.72&46.35&
1.53 \\ 
PG1100+772   &44.78&45.63&45.95&45.93&&46.13&45.85&&46.47& 1.79&46.36&
1.37 \\ 
PG1259+593\tablenotemark{c}   &   \nodata&   \nodata&   \nodata&46.24&&   \nodata&46.13&&   \nodata&   \nodata&46.57&
1.09 \\ 
PG1322+659   &44.54&45.16&45.22&45.21&&45.36&45.16&&45.80& 1.93&45.73&
1.66 \\ 
PG1351+640   &43.30&44.13&45.08&45.11&&45.52&45.71&&45.68& 0.83&45.83&
1.16 \\ 
PG2349-014   &44.71&45.07&45.22&45.20&&45.54&45.30&&45.87& 1.82&45.77&
1.45 \\ 
TON951       &42.95&44.39&44.92&44.90&&45.12&45.05&&45.41& 1.26&45.37&
1.15 \\

\enddata
\tablecomments{Values are the logarithm of luminosity in units of \ergs.
\Lx: 2--0.2\,keV; \Lxfuv: 0.2\,keV--700\AA; \Lfuvo: 700\AA--1\micron; 
\Lir: 1\micron--100\micron; \lbol: 2\,keV--100\micron.}
\tablenotetext{a}{Case A: IR bump is scaled to match \aored\ at 1\micron; 
Case B: IR bump is scaled to match \auvo\ at 1\micron.  
\aored\ is ignored and \auvo\ is used for the entire optical region.  
}
\tablenotetext{b}{R$_A$,R$_B$: ratios of the integral 
\lbol to 9\lllambda(5100\AA) for case A and B, respectively.
}
\tablenotetext{c}{\lbol\ does not include \Lx\ and \Lxfuv.}

\end{deluxetable}

Since we do not use actual measurements in the IR (few exist for our
sample objects), we scale the IR bump
in two ways so that we have a range of the estimated IR luminosity.
In case A, we match the IR bump with the extrapolation of the
red optical power-law of \aored\ at 1\micron;  in case B, 
we ignore \aored\ and use NUV-optical power-law of 
\auvo\ for the entire optical region, and
match the IR bump with the extrapolation of 
\auvo\ at 1\micron.
Therefore, we have two integral luminosities for FUV-optical \Lfuvo\
(700\AA--1\micron), two
estimated luminosities of the IR bump \Lir\ (1--100\micron), and two
estimates of \lbol\ (2\,keV--100\micron).  
These two cases give consistent results for
\Lfuvo, but result in a big difference, up to a factor of 2.5 in \Lir,
when \auvo\ and \aored\ are significantly different.
In any case, \Lir\ is comparable with \Lfuvo\ and contributes
significantly to \lbol.  
Table~\ref{tb:lirluv} shows the comparison between \Lir\ and \Lfuvo.
The distribution of \Lir/\Lfuvo\ has a large dispersion,
in general agreement with the data of \citet{Elvi94}.

\begin{deluxetable}{lccccc}
\tabletypesize{\scriptsize}
%\tabletypesize{\tiny}
%\rotate
\tablecaption{Statistics of \Lir/\Lfuvo \label{tb:lirluv}}
\tablewidth{0pt}
\tablehead{
\colhead{} & 
\colhead{Median} & 
\colhead{Mean} & 
%\colhead{stddev} & 
\colhead{Min} & 
\colhead{Max} 
}
\startdata

%                & Median &Mean &stddev  & min   & max  \\
Case A          & 1.61   &1.76   $\pm$0.63  & 0.78  & 2.78  \\
Case B          & 1.20   &1.49   $\pm$0.91  & 0.70  & 4.00  \\
\citet{Elvi94}\tablenotemark{a}  & 1.17   &1.56   $\pm$1.23  & 0.58  & 5.89  \\

\enddata

\tablenotetext{a}{\Lfuvo\ is between 0.1--1\micron.  Data are obtained
from their Table~15 for 34 objects, for which L(1--10\micron) and
L(10-100\micron) are not given as upper limits. 
}

\end{deluxetable}

Figure~\ref{fg:lbol} compares our integral \lbol\ with the bolometric
luminosity estimated using the empirical formula from a monochromatic
optical luminosity, $\lbol=9\lllambda$(5100\AA).  The ratios of
\lbol/9\lllambda(5100\AA) are listed in Table~\ref{tb:lbol}.  (The
\lllambda(5100\AA) is measured in a local continuum. See \S\ref{sec:hb}).
There exists a strong correlation between the integral \lbol\ and
9\lllambda(5100\AA), but our results are 30--70\% ($\sim 0.1-0.2$~dex)
larger in general and seems to agree better with the correction factor
determined by
\citet{Elvi94}.  These imply that the bolometric correction factor
may be larger than 9, and more like 13 for using
\lllambda(5100\AA).

%\clearpage    
\begin{figure}
\epsscale{.40}
\plotone{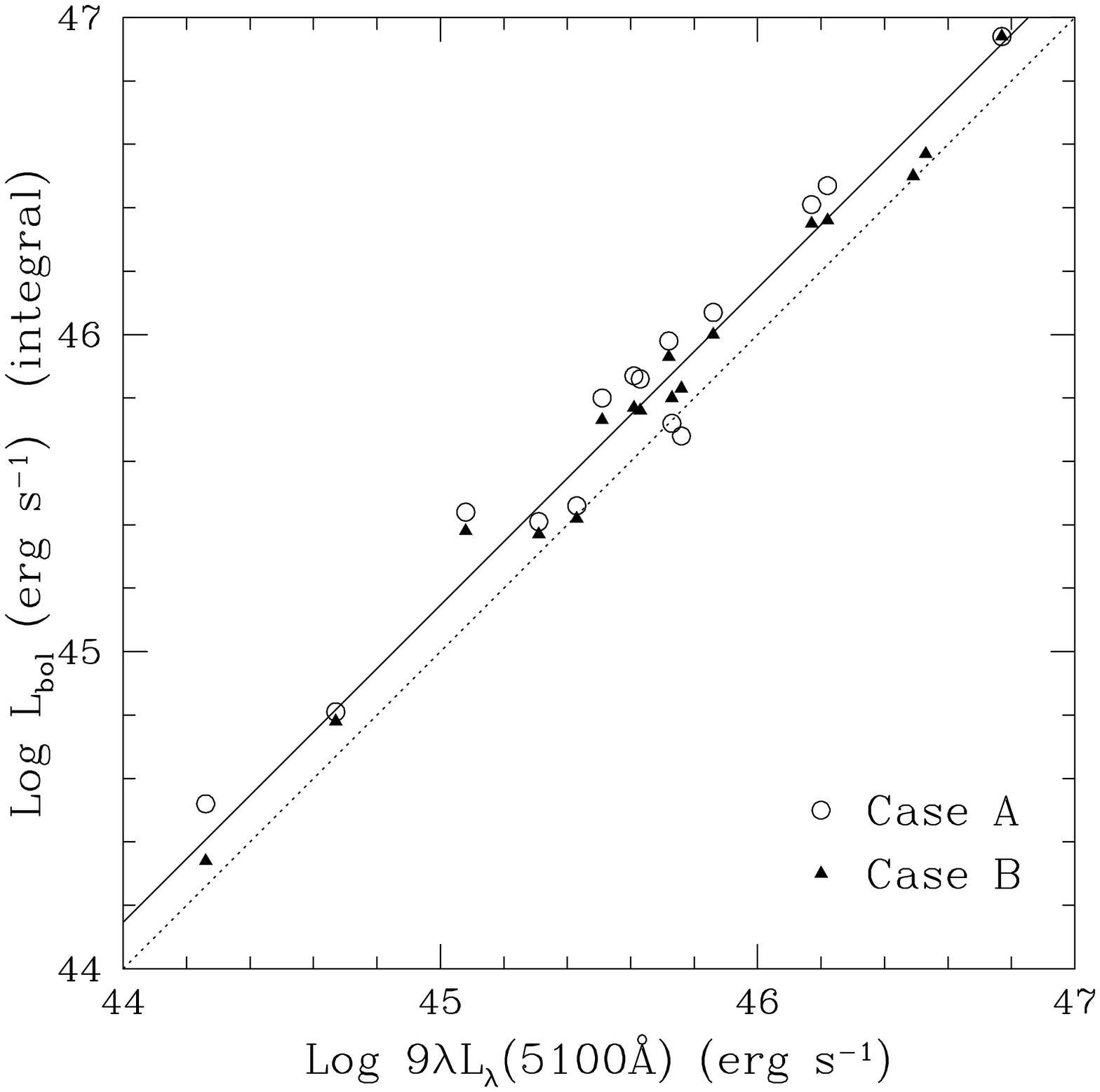}
\caption{Comparison between integral \lbol\ and 9\lllambda(5100\AA).
The solid line indicates $\lbol=1.5 \times 9\lllambda(5100\mbox{\AA})$,
it is not a least square fit.
\label{fg:lbol}} 
\end{figure}   %\clearpage

Although we obtain accurate FUV-to-optical luminosities from our SEDs, a
few sources contribute to the uncertainties of the integral bolometric
luminosity.  First, \citet{Elvi94} warned about the use of mean SEDs,
because the shapes of SEDs in individual objects have large
dispersion.  Second, our scaling of the IR bump may cause large
uncertainties since we use simple extrapolations of the NUV-optical
and red optical
power-laws.  We have seen a difference of 30\% (0.1~dex) in \lbol\
for case A and B.  Third, we have no data for \Lxfuv, and the cutoff
wavelength of 700\AA\ for \Lfuv\ is chosen subjectively.  Finally, the
X-ray data were not obtained simultaneously with our FUV-optical
spectra, and the X-ray variability
is another source of uncertainty.  These should
be kept in mind when interpreting our integral \lbol.  

For simplicity and easy comparison with other work, we will use
\lbol=9\lllambda(5100\AA) for the bolometric luminosity in the rest of
the paper.  This choice does not change any statistical significance
in analyses involving \lbol, and it is easy to compare with our
integral bolometric luminosity by using Table~\ref{tb:lbol} to correct
for individual objects. 

%%%%%%%%%%%%%%%%%%%%%%%%%%%%%%%%%%%%%%%%%%%%%%%%%%%%%%
\subsection{Estimation of Black Hole Mass\label{sec:hb}}

We have estimated the black hole mass and accretion rate using a
recently developed method based on reverberation mapping of the
broad-line region (BLR) and on the assumption of virial motion
\citep{Kasp00},

%and the results of reverberation mapping \citep{Kasp00}.
\begin{equation}
\label{eq:Mbh}
M_{BH} = R_{BLR}\ v^2/G.
\end{equation} 
\hb\ is used for estimating the velocity dispersion,
$v = \sqrt{3}/2$~FWHM(\hb),
and $R_{BLR}$ is the size of the broad line region and can be estimated
empirically from reverberation mapping studies \citep{Kasp00},
\begin{equation}
\label{eq:Rblr}
 R_{BLR} = 32.9^{+2.0}_{-1.9} \left[ \frac{\lambda L_\lambda (5100 \mbox{\AA})}
{10^{44} \mbox{erg s}^{-1} } \right]^{0.70\pm0.033} \mbox{  light days}.
\end{equation}
We use the bolometric luminosity 
$L_{bol} = 9\lambda L_\lambda(5100 \mbox{\AA})$.
Given the black hole mass and bolometric
luminosity, we can also estimate the Eddington ratio, \lledd.

We measure the FWHM(\hb) by fitting the \hb\ region with the IRAF
task {\it specfit} \citep{Kris94}.  A local power-law continuum,
the \oiii\ lines, and \heiiol\
are also fitted together with \hb\ (Fig.~\ref{fg:fithb}).
We use a broad and a narrow Gaussian component to fit the \hb\
broad line, and allow a relative wavelength shift between the two
components to account for the \hb\ asymmetry.  A narrow-line-region
(NLR) component of \hb\ is also introduced, but it is often negligible.
Both the width and the wavelength of this NLR component are tied
with those of \oiiil.  The intensity ratio of \oiiil/$\lambda$4959
is assumed to be 3:1 based on their statistical weights.  A single
Gaussian profile is fitted for each of \oiiill.

%\clearpage    
\begin{figure}
\epsscale{0.8}
\plotone{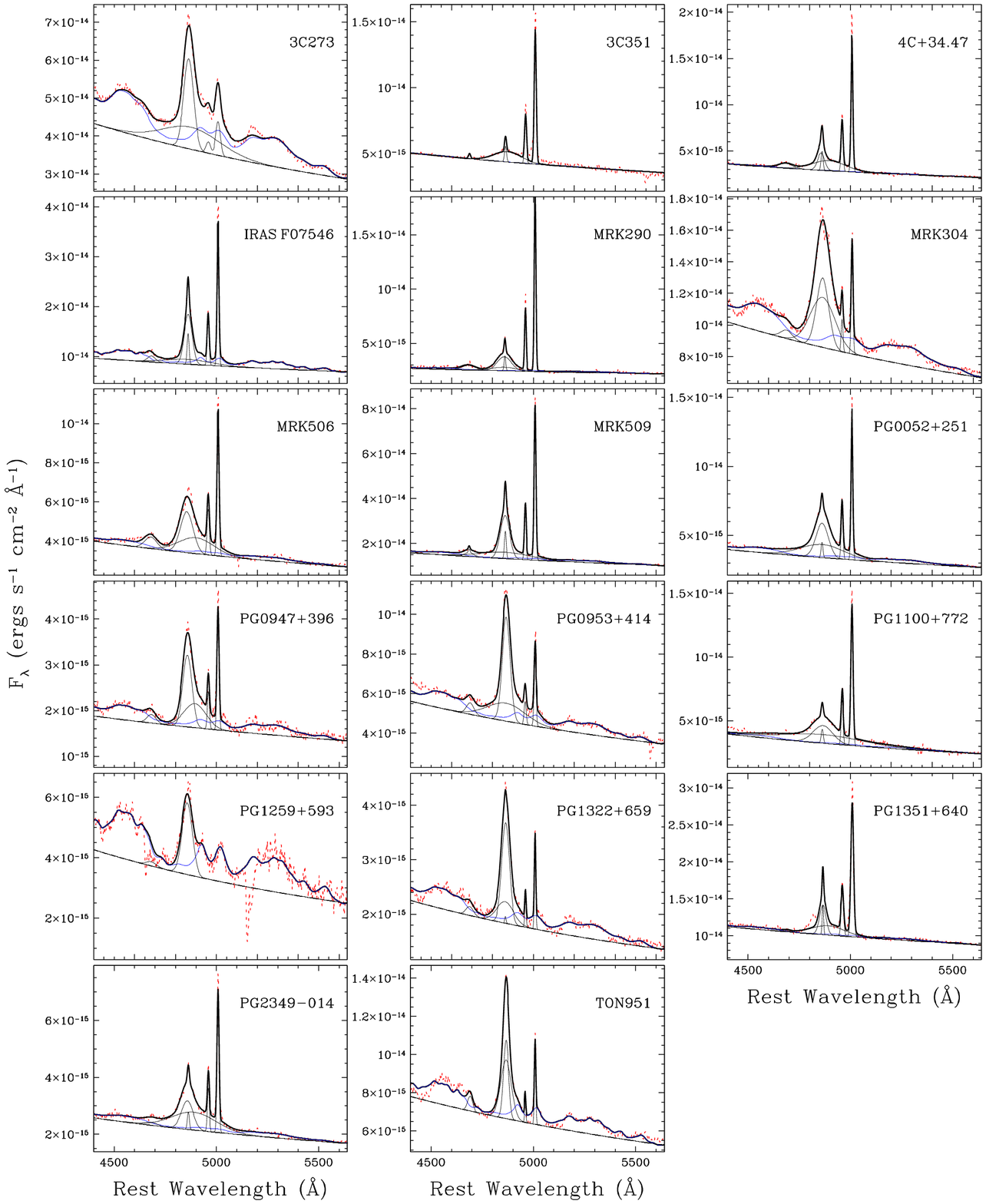}
\caption{Model fitting to the \hb\ region.  The dotted lines show
the data, and the thick solid lines the fitting results.  Also
shown are the local continua, \feii\ template, and individual
components of \hb, \oiii,  and \heiiol.
\label{fg:fithb}} 
\end{figure}   %\clearpage

Broad \feii\ emission blends are often strong in this region,
especially when \oiii\ is weak.  In order to remove the \feii\
contamination, we use an \feii\ template from \citet{BG92}.  The
strength of the \feii\ template is free to vary in the fitting process,
and it is also broadened to be consistent with 
FWHM(\hb) for each object.

We calculate the final FWHM(\hb) from the fitted \hb\ model profiles.
Specifically, we exclude the NLR component, and use the other two
fitted Gaussian components, taking into account the relative
wavelength shift between these two components.  The uncertainty of
FWHM(\hb) is less than 10\%.

The rest-frame reference is defined with \oiii$\lambda$5006.8, and the
fitted wavelength of \oiiil\ is used to calculate the redshift.  The
uncertainty of the redshift is $<0.0002$ for all objects except for
PG1259+592, which has very weak \oiii, and has a redshift uncertainty
of 0.002.

Table~\ref{tb:mbh} lists the 
fitting results, together with calculated \mbh\ and \lledd\ etc. 
Six objects
in our sample also have black hole masses derived from reverberation mapping
studies (Kaspi et al 2000).  They are also listed in the table.
Our results are usually larger,
because we have excluded the \hb\ NLR component when obtaining
the FWHM(\hb).  Therefore we have a larger FWHM(\hb) and hence a
larger \mbh.  As was also noticed by \citet{Boro02} and \citet{Vest02},
3C351 (i.e., PG~1704+608) is an extreme case, in which
the very narrow \hb\ component on top of the broad \hb\ emission
is identified as an NLR component with the same width as \oiii\ (690\,\kms).
It is excluded from calculating FWHM(\hb) in our study, but not in
\citet{Kasp00}\footnote{Their measured mean FWHM(\hb) is 890\,\kms, and it is
400\,\kms\ from rms spectrum.  The rms spectra in the reverberation
mapping studies are not completely
free of constant narrow components \citep{Pete98}.}.
This results in a huge difference in FWHM(\hb)
and hence in \mbh.

Another uncertainty of the FWHM(\hb) comes from the uncertainty
of the fitted local continuum level.  Assuming a single Gaussian
profile, if the continuum is lowered by 10\% of the line peak,
the estimated FWHM will increase by $\sim$ 7\%, and the calculated
black hole mass from FWHM(\hb) and the continuum luminosity based on
Eq.\ref{eq:Mbh} and \ref{eq:Rblr} will increase by 7\%.

% On the other hand, sometimes we had to scale the optical spectra to
% match the \hst\ NUV spectra when we combined them
% (\S\ref{sec:construction}).  If, however, the flux calibration of
% \hst\ spectra is wrong, for example, a scaling factor of 2 (very
% large) for the optical spectra can change the estimated black hole
% mass by 60\% based on Eq.~\ref{eq:Mbh} and \ref{eq:Rblr}.  

On the other hand, sometimes we had to scale the optical spectra to
match the \hst\ NUV spectra due to likely source variability.  A
scaling of 30\% can change the estimated black hole mass by 20\%.  The
AGN intrinsic variability can also change the emission-line profile,
and hence the estimated black hole mass.  Different approaches for
spectral measurements in different studies can lead to large
discrepancies in reported masses.  Due to all the above reasons, the
estimated black hole mass from different studies can easily differ by
a factor of a few, in our case, a maximum factor of 5 for 3C273
(excluding the extreme case of 3C351).  In fact, \citet{Vest02}
compared the black hole masses estimated from reverberation mapping
and from single-epoch optical observation, and concluded that they
agree within factors of 3, 6, and 10 with probabilities of 80\%, 90\%,
and 95\%, respectively.
However, if one keeps
consistency in measurements and calculation for a sample, the relative
uncertainty in the estimated black hole masses within the sample
should be much smaller.

\begin{deluxetable}{lccccrcl}
\tabletypesize{\scriptsize}
%\tabletypesize{\tiny}
%\rotate
\tablecaption{Black Hole Mass and Eddington Ratio \label{tb:mbh}}
\tablewidth{0pt}
\tablehead{
\colhead{Object} &
\colhead{z} &
\colhead{FWHM(\hb)} &
\colhead{\flambda(5100\AA)\tablenotemark{a}} &
\colhead{\lllambda(5100\AA)\tablenotemark{b}} &
\colhead{\mbh} &
\colhead{\lledd} &
\colhead{\mbh(rev)\tablenotemark{c}}  \\
\colhead{} &
\colhead{} &
\colhead{(\kms)} &
\colhead{(\fluxl)} &
\colhead{log(erg s$^{-1}$)} &
\colhead{($10^8M\sun$)} &
\colhead{} &
\colhead{($10^8M\sun$)} 
}
\startdata
%  obj          &      z &hbfwhm&   F5100&LL5100 &   m8 &lledd \\
  3C273        & 0.1576 &4115  &3.41E-14& 45.82 &11.50 & 0.41 &  2.35\\
  3C351        & 0.3730 &9760  &4.11E-15& 45.54 &41.18 & 0.06 &  0.075\\
  4C+34.47     & 0.2055 &3520  &2.64E-15& 44.91 & 1.94 & 0.30 &  \\
  IR07546+3928 & 0.0953 &2965  &7.89E-15& 44.78 & 1.12 & 0.39 &  \\
  MRK290       & 0.0303 &5505  &2.38E-15& 43.31 & 0.36 & 0.04 &  \\
  MRK304       & 0.0657 &5570  &7.96E-15& 44.48 & 2.43 & 0.09 &  \\
  MRK506       & 0.0428 &5520  &3.15E-15& 43.72 & 0.70 & 0.05 &  \\
  MRK509       & 0.0345 &3630  &1.22E-14& 44.13 & 0.59 & 0.16 &  0.92 \\
  PG0052+251   & 0.1544 &5465  &3.12E-15& 44.77 & 3.73 & 0.11 &  3.02\\
  PG0947+396   & 0.2057 &3810  &1.53E-15& 44.68 & 1.57 & 0.22 &  \\
  PG0953+414   & 0.2338 &3155  &4.23E-15& 45.22 & 2.57 & 0.46 &  1.64\\
  PG1100+772   & 0.3114 &9300  &2.94E-15& 45.27 &24.20 & 0.06 &  \\
  PG1259+593   & 0.4769 &3615  &3.11E-15& 45.58 & 6.03 & 0.45 &  \\
  PG1322+659   & 0.1684 &3030  &1.67E-15& 44.56 & 0.82 & 0.32 &  \\
  PG1351+640   & 0.0882 &2840  &9.69E-15& 44.81 & 1.08 & 0.43 &  0.30\\
  PG2349$-$014   & 0.1740 &5900  &1.98E-15& 44.66 & 3.64 & 0.09 &  \\
  TON951       & 0.0643 &2390  &6.19E-15& 44.36 & 0.37 & 0.45 & 
\enddata

%\tablecomments{}
\tablenotetext{a}{Rest-frame flux density at 5100\AA.}
\tablenotetext{b}{Assuming zero cosmological constant, \Hoeq, 
and $q_0=0.5$, same as
the cosmology used in \citet{Kasp00}.}
\tablenotetext{c}{Black hole mass measured from reverberation mapping
studies
\citep{Kasp00}.}

\end{deluxetable}

\subsection{Correlation Analyses\label{sec:corr}}

It is natural to think that the properties of the UV bump, thought to
arise from the accretion disk, are governed by the AGN fundamental
parameters, such as \mbh\ and \lledd.  For example, standard thin disk
models predict low disk temperatures for high \mbh\ and/or low \lledd\
\citep{ShaSun73} and therefore longer break wavelengths.  We looked
for such correlations in our sample, but we do not find any significant
correlation between the UV bump properties we have measured (\afuv,
\auvoa, \deltaa, and \lbreak) and \mbh, or \lledd.
Table~\ref{tb:corr} lists the Pearson correlation coefficients for
selected parameters.  A principal component analysis has also been
performed on these parameters, and no hidden correlations are
revealed.  We also group the objects into subsamples based on \lledd\
(or \mbh), but there is still no evidence of a correlation within
subsamples (see Fig.~\ref{fg:fuvambh} for an example).  However,
\citet{Scot04} found a weak correlation at 96\% confidence level
between \afuv\ and \mbh\ in their subsample of 21 AGNs.  There are
only 4 objects in common between these two studies. The inconsistency
between the results most likely arises from the small size of the
samples.  In any case, the fact that there are no or weak correlations
suggests that \mbh\ and \lledd\ are not the only parameters underlying
the observed properties of the UV bump.  Other factors, such as the
disk inclination, intrinsic reddening or other unidentified parameters
must also play an important role (\S\ref{sec:model}).  We also note
that a large spread over \mbh\ and \lledd\ within the small samples
may wash out any correlation with spectral index, but our sample is
too small to address this.

\begin{deluxetable}{rrrrrrrrr}
\tabletypesize{\scriptsize}
%\rotate
\tablecaption{Pearson Correlation Coefficients ($r$)\label{tb:corr}}
\tablewidth{0pt}
\tablehead{
\colhead{} & 
\colhead{log(\mbh)} & 
\colhead{\lledd} & 
\colhead{\afuv} & 
\colhead{\auvo} & 
\colhead{\deltaa} & 
\colhead{\lbreak} & 
\colhead{\ebv} & 
\colhead{\ax}
}
\startdata
% %             1      2      3      4      5      6      7	\\
% %       logmbh  lledd   fuva  diffawavetur    Ebv     ax	\\
% log(\mbh) & 1.00	\\
%   \lledd &$-$0.16 & 1.00	\\
%    \afuv &$-$0.23 &$-$0.25 & 1.00	\\
%\auvoa
%   \deltaa &$-$0.04 &$-$0.34 & {\bf 0.82} & 1.00	\\
% \lbreak &0.04 & 0.27 &$-$0.06 &$-$0.26 & 1.00	\\
%     \ebv &$-$0.16 &$-$0.25 & {\bf 0.51} & 0.36 & 0.15 & 1.00	\\
%      \ax & 0.23 &$-$0.24 &$-$0.16 & 0.04 &$-$0.39 &$-${\bf 0.64} & 1.00

%       #logmbh  lledd   fuva   uvoa  diffawavebre    Ebv     ax
 log(\mbh)& 1.00 \\
   \lledd &$-$0.16 & 1.00 \\
    \afuv & 0.23 &$-$0.06 & 1.00 \\
  \auvoa  &$-$0.33 & 0.12 & 0.46 & 1.00 \\
  \deltaa & 0.47 &$-$0.14 & {\bf 0.82} &$-$0.13 & 1.00 \\
  \lbreak & 0.02 & 0.06 & 0.34 & 0.39 & 0.14 & 1.00 \\
     \ebv &$-$0.16 &$-$0.24 & 0.31 & 0.30 & 0.15 & 0.28 & 1.00 \\
      \ax & 0.23 &$-$0.24 &$-$0.09 &$-$0.31 & 0.15 & {\bf $-$0.67} &
		{\bf $-$0.64} & 1.00
\enddata 

\tablecomments{17 objects are used except for \ax\ (15 objects)
and \lbreak\ (16 objects).  The
chance probability ($p$) of 1\%, 2\%, and 5\% corresponds to a correlation
coefficient of 0.61, 0.56, and 0.48, respectively, for 17 objects.  Large
correlation coefficients marked in bold face do not reveal significant
physical correlations (see \S\ref{sec:corr} for detail).}

\end{deluxetable}

%\clearpage    
\begin{figure}
\epsscale{.50}
\plotone{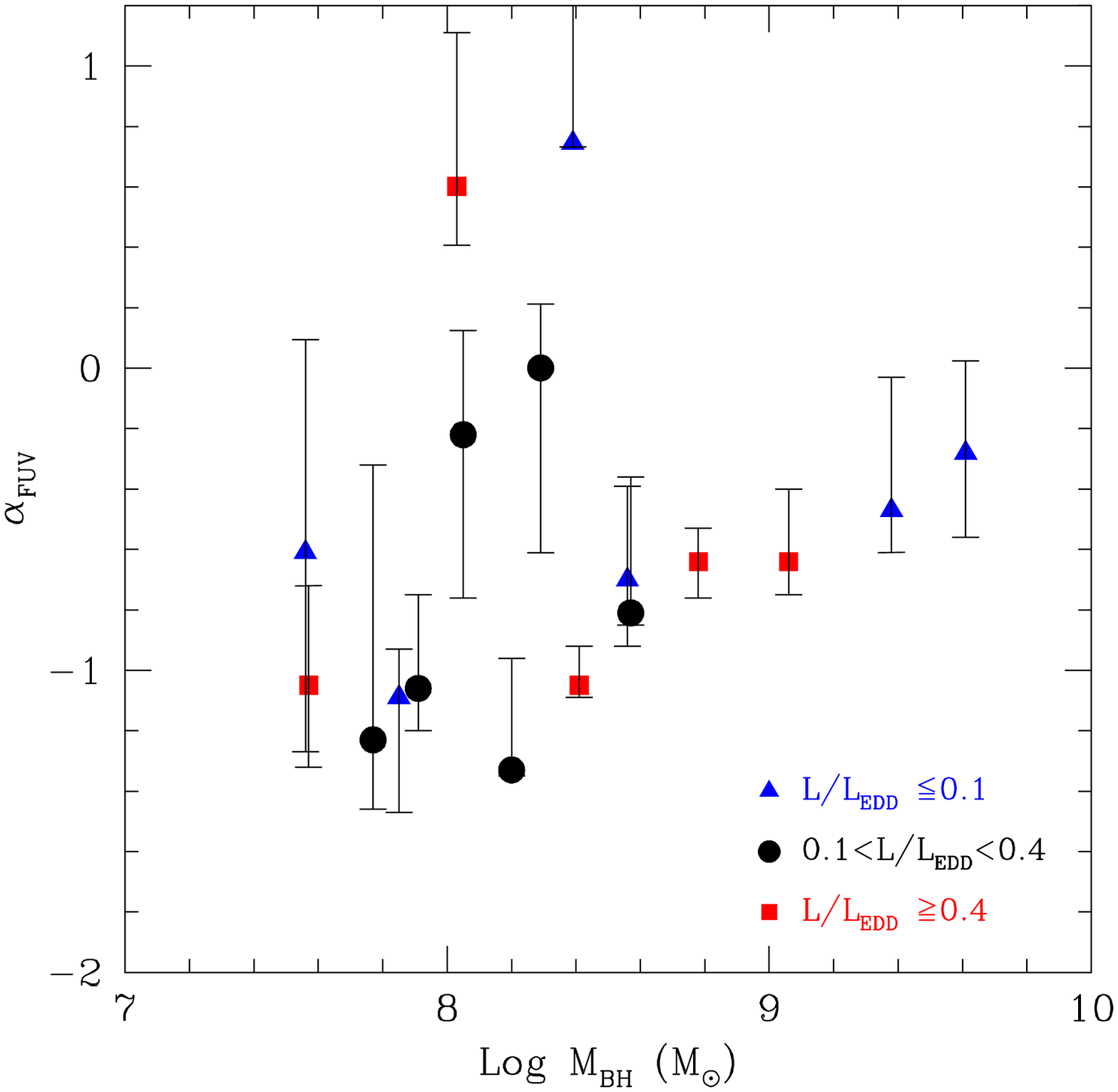}
\caption{No correlation between \afuv\ and \mbh\ for the total
sample or for subsamples grouped by \lledd.
\label{fg:fuvambh}}
\end{figure}   %\clearpage 

A strong correlation ($r=0.82, p=0.0001$) seen in Table~\ref{tb:corr}
is between \deltaa\ ($=\afuv-\auvoa$) and \afuv.  This is simply
because the distribution of the NUV-optical spectra index \auvoa\
is relatively narrow, while \afuv\ spans a wide range.
Figure~\ref{fg:ahist} shows this clearly.  Most objects have \auvoa\
between $-2$ and $-1$ with a median value of $-1.52$,
but the distribution of \afuv\  is broader (Fig.~\ref{fg:ahist}b).

There seems to be an anti-correlation between \ax\ and \lbreak, but
there are only 14 objects with values of both \ax\ and \lbreak\ ($r=-0.67,
p=0.01$).   Since \lbreak\ has very large uncertainty, this
correlation should not be treated seriously.

% There exists a marginally significant correlation ($r=0.51, p=0.04$)
% between \afuv\ and \ebv.  It may indicate that the Galactic
% reddening is systematically under-corrected for in our FUV spectra.
% However, when constructing a \fuse\ FUV composite spectrum from a
% larger AGN sample, of which our sample is a subsample,
% \citet{Scot04} find a marginally significant correlation between
% their $\alpha_\nu (FUV)$ and \ebv, indicating a possible {\it
% over-correction} for Galactic reddening.  Therefore, we argue that
% the correlation we see here is likely due to chance.

We also see a correlation ($r=0.64, p=0.01$, for 15 objects) between
\ebv\ and \ax, but this is largely due to an outlier, IRAS~F07546+3928, 
with the softest $\ax=-2.16$.  Without this outlier, the correlation
disappears ($p=0.23$).  After careful checking, we find no other
correlations that are created or destroyed by outliers.

%\clearpage    
\begin{figure}
%\epsscale{.50}
\includegraphics[angle=270,scale=0.25]{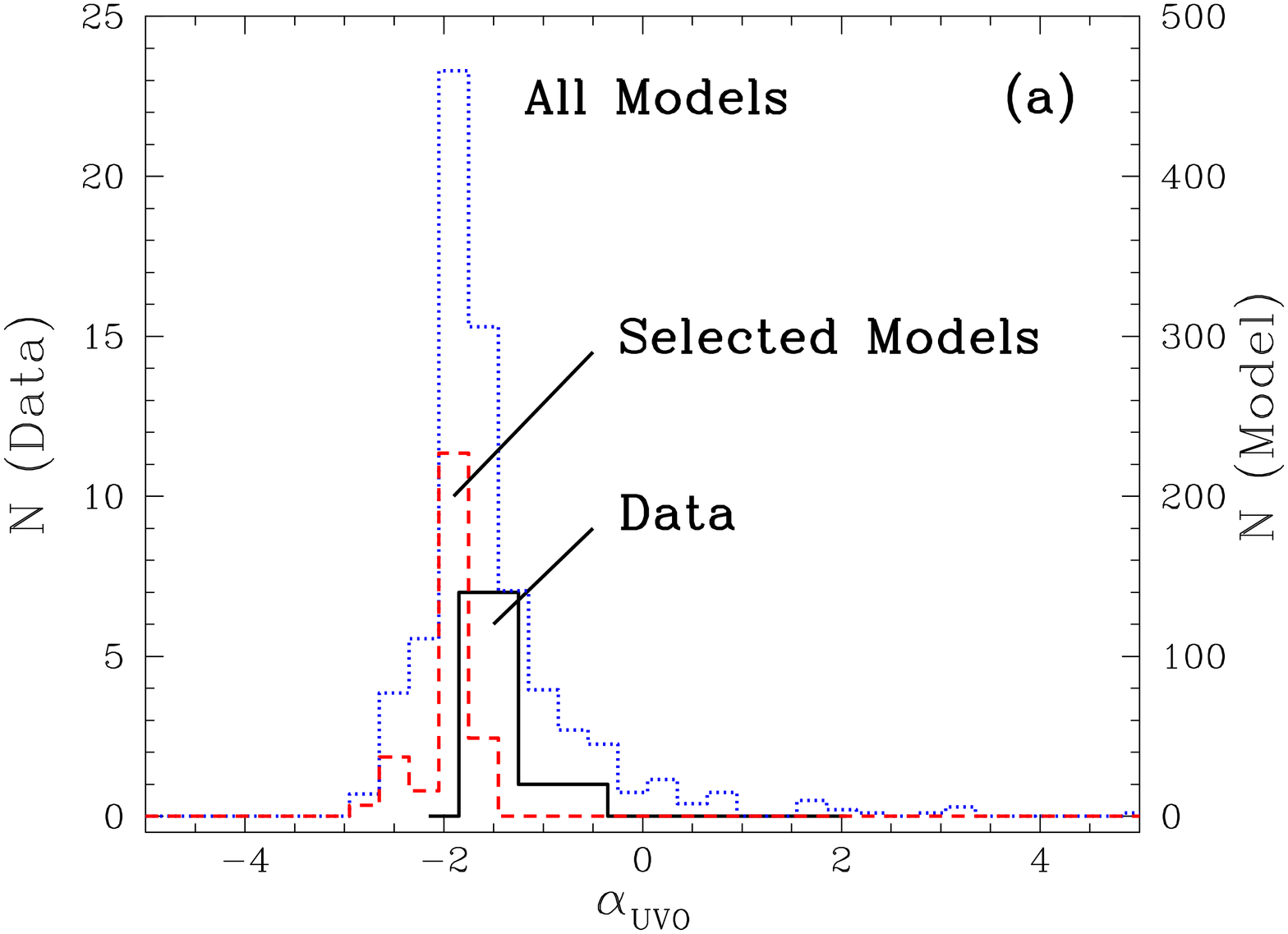}
~ ~ ~
\includegraphics[angle=270,scale=0.25]{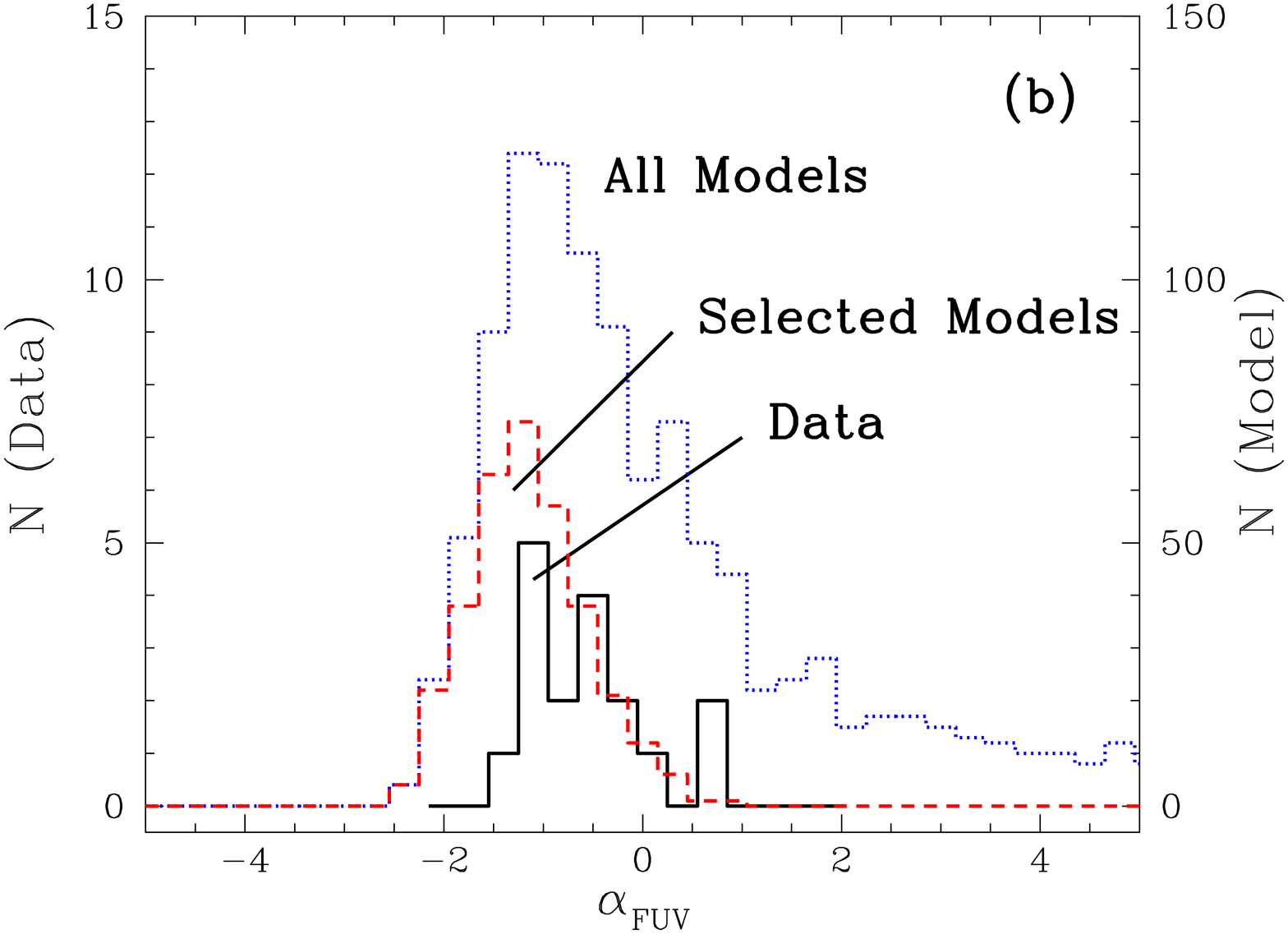}
\caption{Distributions of \auvoa\ and \afuv\ 
of the sample (solid line), all models (dotted line,
\S\ref{sec:model})
and the selected models (dashed line, $10^8\msun < \mbh \leq 4\times
10^9\msun$, $0.03< \lledd < 0.3$). 
Note that no selected models
are as red as $\auvo>-1$.
Two reddest objects ($\auvo>-1$) are \irasobj\ and PG~1351+640,
both of which show evidence of intrinsic dust reddening.
\label{fg:ahist}}
\end{figure}   %\clearpage

\section{COMPARISON WITH THIN-DISK MODELS\label{sec:model}}

As mentioned above, the expected correlations between the UV bump and
\mbh\ or \lledd\ may be mitigated by the small sample size, a large
variation in these parameters within the sample, and by other
parameters affecting the spectrum of the accretion disk.  We used the
thin disk model developed by \citet{Hube00} to investigate this.

The models are constructed for a non-LTE disk with 5 free parameters:
black hole mass, mass accretion rate (\mdot), viscosity parameter
$\alphav$, black hole spin, and inclination angle $\cos i$.  The total
spectrum of a disk is integrated over the individual annuli, taking
into account the inclination angle.  We have chosen a grid of models
with a maximally rotating Kerr black hole with possible values of
\mbh\ between $0.125 \times 10^9$ and $32\times 10^9 \msun$, \mdot\
between $2^{-12}\msunyr$ and $64\msunyr$, $\alphav = 0.01$ or 0.1, and
$\cos i$ between 0.01 and 0.99.  The maximum Eddington ratio is
limited to $\lledd\approx 0.3$.  Above this value, the model disk
becomes geometrically thick and thus no longer self-consistent.

%1386 models

In order to make statistical comparisons between theory and observation, 
we have measured \afuv, \auvoa, and \lbreak\ of the models in the same 
way we have measured them in our data.  Figure~\ref{fg:modelfit}
shows examples of how we measure the model spectra.  The Lyman break
is prominent in some models, especially when
\lledd\ is small.  However, due to relativistic boosting
and aberration, small \cosi\ (edge-on) tends to
smear out the edges and also shift them to shorter wavelengths.
Therefore, to measure \afuv\ for models with different \cosi,
we fit a power-law to a different smooth region of $\sim$200\AA\
blueward of the Lyman break.  We note that this is not exactly the same as
in measuring our data, but this characterizes the UV bump of the
models very well except for those with very strong Lyman edges
(Fig.~\ref{fg:modelfit}), and the estimate is consistent for all
models with and without a strong Lyman edge.  The NUV-optical
spectral index is measured the same way as in our real data by
fitting a power-law to two continuum windows around 1350\AA\
and 5630\AA\ (continuum windows b and d in Table~\ref{tb:index}).
\lbreak\ is calculated from \afuv\ and \auvoa.

%data fitting: Sample="1348-1358,5600-5648"            # 2, 4

%wrange[1]="364-564"
%wrange[2]="427-627"
%wrange[3]="498-698"
%wrange[4]="549-749"
%wrange[5]="576-776"
%wrange[6]="633-833"
%wrange[7]="725-925"
%wrangeuvo="1350-5630"

%\clearpage    
\begin{figure}
\epsscale{0.6}
\plotone{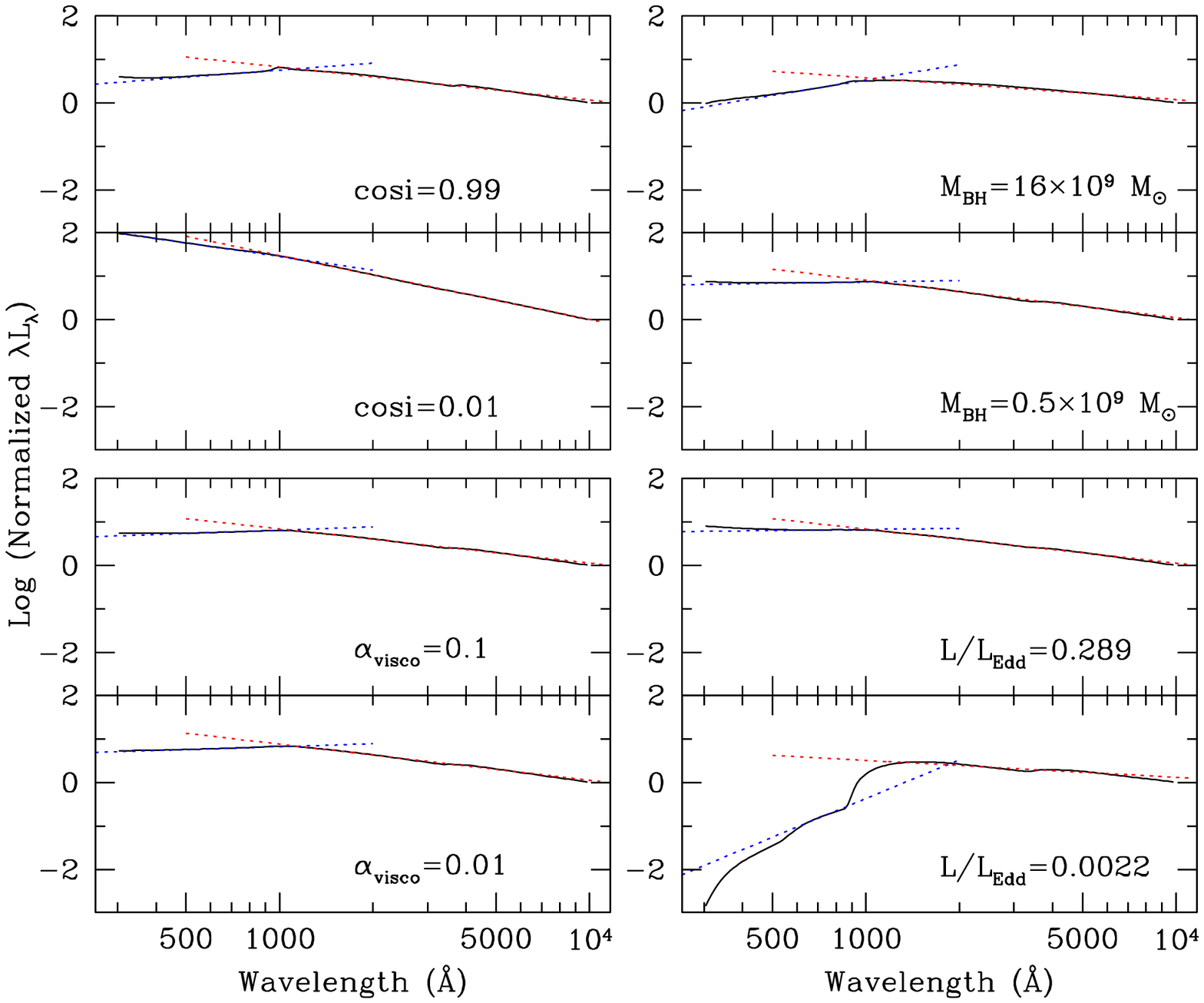}
\caption{Examples of power-law fitting to the disk model spectra.
Solid lines are the model spectra, and dotted lines are the fitted power-laws.
Unless marked in each panel, the default model parameters are: 
$\mbh=1\times 10^9 \msun$, $\lledd=0.145$, $\alphav=0.1$, and $\cosi=0.8$ 
($i=37\degr$, close to face-on).  Note that FUV slopes obviously change
with \mbh, \lledd\ and \cosi\ (but not \alphav), but
UV-optical slopes do not change much.  Also note the strong Lyman break
in the spectrum with low \lledd\ (0.0022).
\label{fg:modelfit}}
\end{figure}   %\clearpage 

Similar to our data, the distribution of \auvo\ measured from all the
models is relatively narrow with a median value of $-1.74$.  We
further select only models with similar black hole masses ($\mbh\sim
10^8-4\times 10^9 \msun$) and Eddington ratios ($\lledd \sim
0.03-0.3$) to those of our sample and compare them with our data
(Fig.~\ref{fg:ahist}a).  The \auvo\ of the selected models has a
median value of $-1.91$ (standard deviation $\sigma=0.26$), roughly in
agreement with our data (median $\auvo=-1.52$, $\sigma=0.35$).  
While there are no selected models that are as red as $\auvo>-1$
(Fig.~\ref{fg:ahist}a), two objects, \irasobj\ and
PG~1351+640, are redder than $\auvo>-1$, but both show obvious
evidence of dust reddening.  
We
note that 7 objects in our sample have $\lledd>0.3$, and 5 objects
have $\mbh<10^8\msun$, and these values of the parameters have not
been covered by our current models.  
Examining the model trends in Figure~3 of Blaes (2004), extrapolating
$L/L_{Edd}$ of the models to 0.5 would probably not change
$\alpha_{UVO}$ by more than 0.1.  However, lower black hole masses
result in smaller (bluer) $\alpha_{UVO}$ in the models, and such
values are not seen in the data.  On the other hand, we have assumed a
near maximal black hole spin in all the models used here.  Lower black
hole spins generally increase $\alpha_{UVO}$ in the models, and would
improve agreement with the data.  It would therefore be worth
exploring such low spin models in the future.

% Acoording to Figure~3 in
% \citet{Blae04}, while low black hole mass would result in a smaller
% (model) \auvo\ and does not favor the agreement between our data and
% the models, extrapolating \lledd\ of the models to 0.5 cannot
% change \auvo\ by more than 0.1.  And on the other hand, we have assumed the
% maximum black hole spin for the current models, but a lower black hole
% spin will favor the agreement between our data and the models in
% \auvo, as can be seen in the same figure mentioned above.

\afuv\ from the models still has a broad distribution, also in
agreement with our data (Fig.~\ref{fg:ahist}b).  Since the
distribution of \auvo\ is narrow, the spectral breaks at the UV bump
(\lbreak\ and $\deltaa\ =\afuv-\auvo$) are mainly defined by the
change of FUV spectral index \afuv\ in both our data and the models.
\afuv\ is sensitive to \mbh, \lledd, and \cosi, but not to the
viscosity parameter $\alphav$, as shown in Figure~\ref{fg:modelfit}.

We further compare our data with the models by showing the changes
of \afuv, \deltaa\ and \lbreak\ with \mbh\ and \lledd\ 
(Fig.~\ref{fg:datamodmbh} and \ref{fg:datamodlledd}). 
We chose the models for extreme face-on (\cosi=0.99)
and extreme edge-on (\cosi=0.01) cases with $\alphav=0.1$.

%\clearpage    
\begin{figure}
\epsscale{1.0}
\plotone{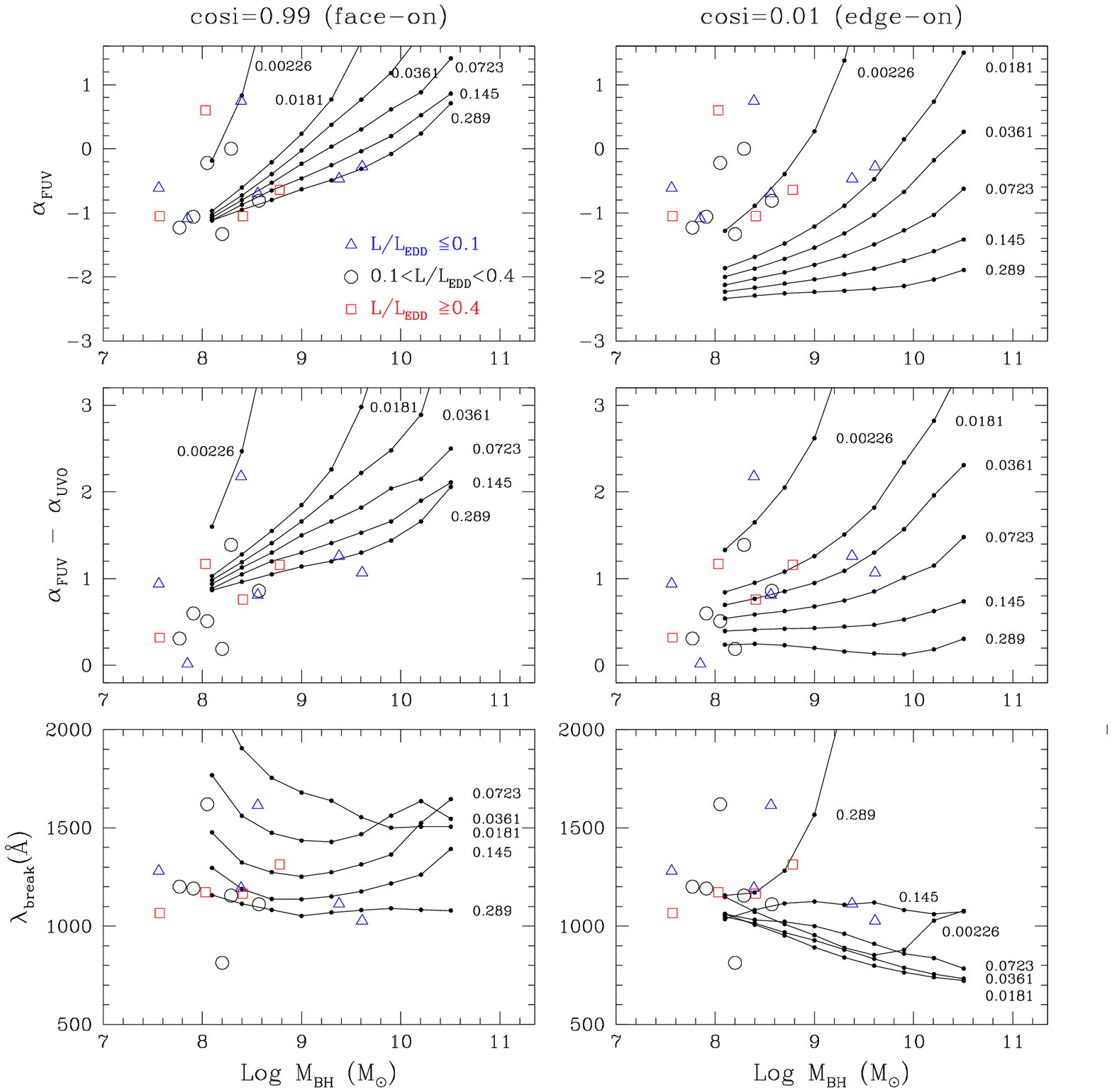}
\caption{Properties of the spectral break vs. black hole mass for
both data and model predictions (face-on and edge-on).  The different 
open symbols are for the individual objects in different \lledd\ 
ranges.  The solid lines are the model predictions.  \lledd\ for 
each line is marked with a number close to the lines.
The irregular patterns (intersections) of the models in the lower panels 
are due to the uncertainty of calculated \lbreak, which is mainly from the way
we measure \afuv\ in models with strong Lyman edge
(see text).
\label{fg:datamodmbh}}
\end{figure} 

%\clearpage    
\begin{figure}
\epsscale{1.0}
\plotone{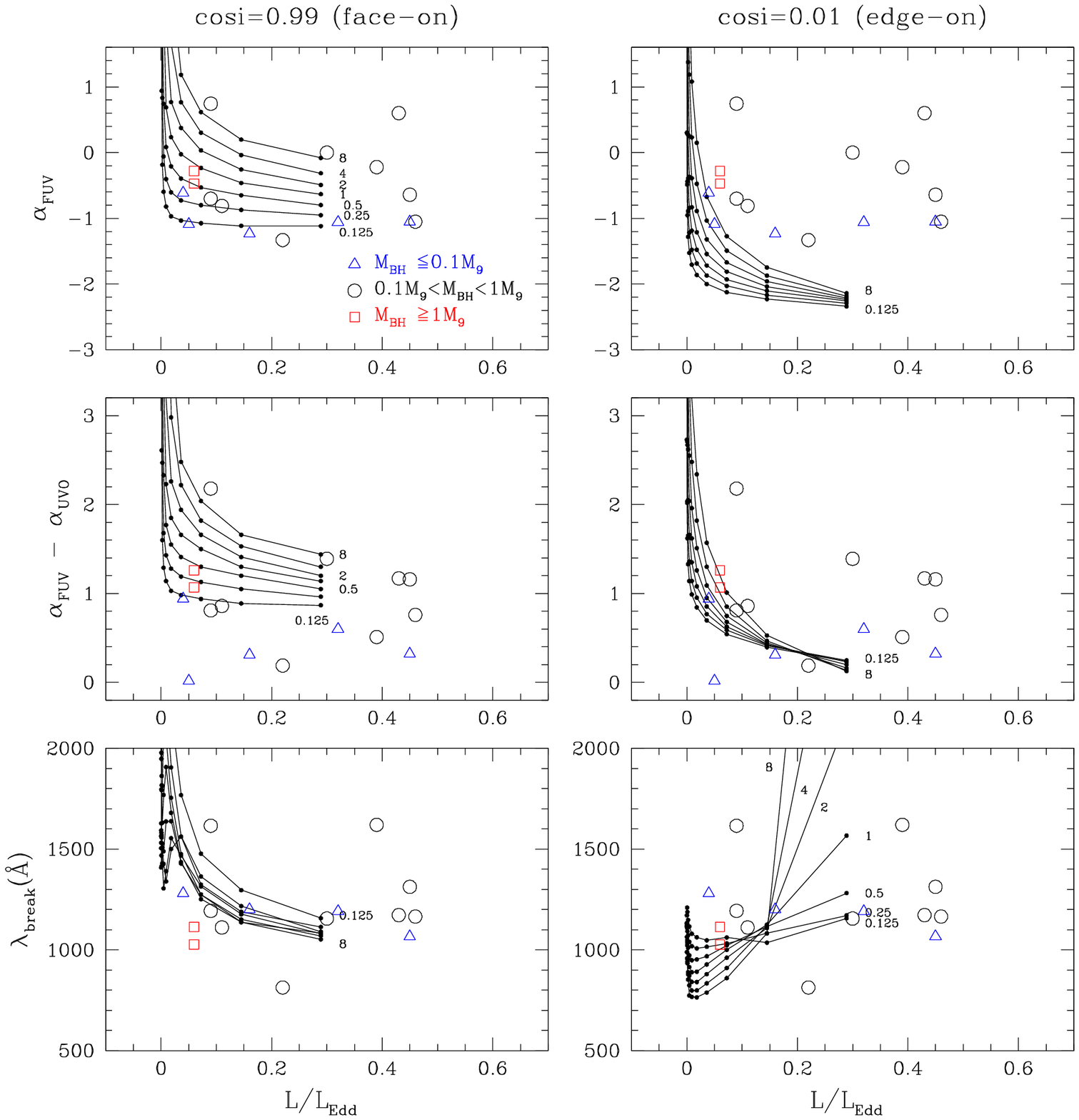}
\caption{
Properties of the spectral break vs. Eddington ratio for
both data and model predictions (face-on and edge-on).  The different
open symbols are for the individual
objects in different \mbh\ ranges.  The solid lines are the model
predictions.  \mbh\ for each line 
is marked with a number close to the lines in units of $M_9$
($M_9=10^9\msun$).
\label{fg:datamodlledd}}
\end{figure}   %\clearpage 

In the face-on models, there is a clear correlation between
\afuv\ (hence \deltaa) and \mbh\ for $\lledd \approx 0.01-0.3$
(Fig.~\ref{fg:datamodmbh}, top-left).  However, no evidence of such
a trend can be seen in our data, even when we group the objects into
subsamples of narrower \lledd\ ranges.  Moreover, the models do not
cover the \afuv-\mbh\ space with enough overlap of the data except
for extremely small \lledd, but these small \lledd\ are not seen
for for our AGNs.  Some
AGNs seems to follow the model prediction in one plot (2-dimensional
space), but they do not match the same model prediction in other
plots (other dimensions).  Although we do not have information on
the inclinations of objects in our sample, large disk inclinations 
only increase the discrepancy (Fig.~\ref{fg:datamodmbh}, top-right).  
We will discuss this inconsistency more in \S\ref{sec:disc}.

%\clearpage    
\begin{figure}[h]
%\begin{figure}
%\epsscale{.50}
\includegraphics[angle=270,scale=0.25]{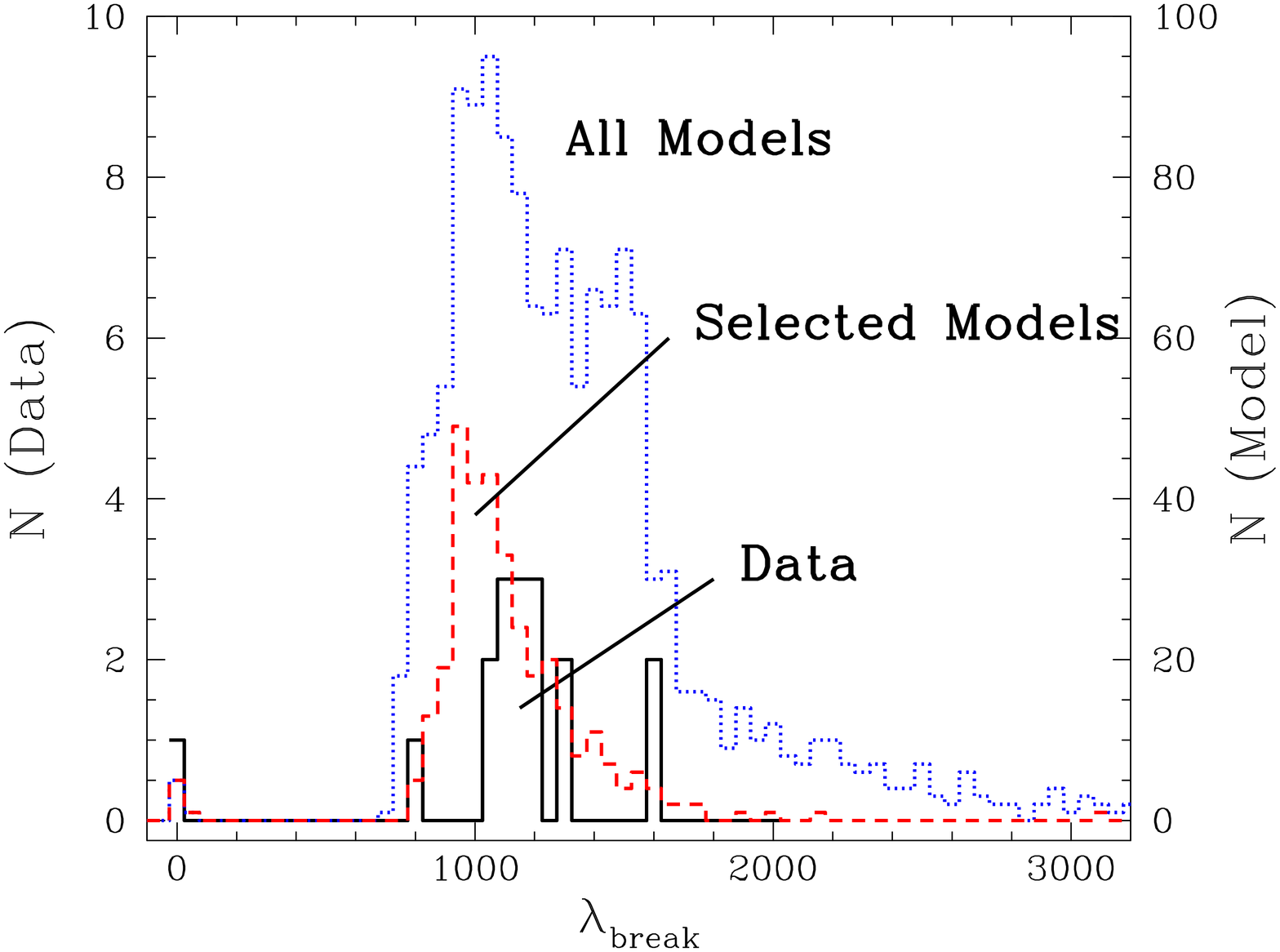}
\caption{Distribution of \lbreak\ of the sample (solid line), all
models (dotted line), and the selected models (dashed line, $10^8\msun
< \mbh \leq 4\times
10^9\msun$, $0.03< \lledd < 0.3$).
Note the peak for the models is around 1000\AA.
The second bump between 1200--1400\AA\ for all models 
is due to an overestimate for
models with a strong Lyman edge. 
\label{fg:mlbreakhist}}
\end{figure}   %\clearpage

\lbreak\ does not seem to correlate with \mbh\
(Fig.~\ref{fg:datamodmbh}, bottom-right).  For large \lledd\ in
face-on cases, \lbreak\ is close to 1000\AA, but for smaller \lledd,
\lbreak\ goes to longer wavelengths.  This is because \afuv\ is
underestimated for models with a strong Lyman edge, as can be seen in
Figure~\ref{fg:modelfit} (bottom-right).  The distribution of model
\lbreak\ also shows this clearly (Fig.~\ref{fg:mlbreakhist}).  While
there is a peak around 1000\AA\ in the histogram, there is also a
second bump between 1200--1400\AA\ which accounts for the models with
a strong Lyman edge.  This bump disappears in the case of the selected
models, where models with small \lledd\ (strong Lyman edge) are
excluded.  In the edge-on cases (Fig.~\ref{fg:datamodmbh},
bottom-right), large inclination causes strong relativistic effects
which smear out the Lyman edge, bringing the measured \lbreak\ back to
around 1000\AA.  The extremely large values of \lbreak\ for
\lledd=0.289 simply indicate that there is not a clear spectral break,
because $\afuv \approx \auvo$.

We have also plotted \afuv, \deltaa, and \lbreak\ against \lledd\ for
different \mbh\ (Fig.~\ref{fg:datamodlledd}).  No clear correlation
is seen.  If we could extend \lledd\ of the models to above 0.3,
the models seem to cover a region that overlaps with our data points
(face-on), but the \mbh\ required for the models needs to extend
above $10^9 \msun$, much higher
than those calculated for most of our AGNs.
The apparent correlation seen in \lbreak-\lledd\ is largely biased
by models with small \lledd, for which \lbreak\ is overestimated
due to the strong Lyman edge.

Large inclination angle (small \cosi, edge-on) has several effects 
on the models:
(1) it decreases (flattens) \afuv, weakening the correlation
between \afuv\ and \mbh\ seen in face-on models 
(Fig.~\ref{fg:datamodmbh}, top-left);
(2) it results in smaller difference between \afuv\ and \auvo\, 
and tends to smear out the UV bump ($\afuv \approx \auvo$),
resulting in unrealistically large (or small) \lbreak\ 
(Fig.~\ref{fg:datamodmbh}, bottom-right);
(3) it forces the models to a narrow region in \afuv-\lledd\ and
\deltaa-\lledd\ space regardless of \mbh\ (Fig.~\ref{fg:datamodlledd}).

\section{DISCUSSION\label{sec:disc}}

%for edge on, parameter space becomes narrow, (physical reason?)

Our data set of quasi-simultaneous FUV-to-optical spectrophotometry is
the first of its kind.  These spectra, with their FUV coverage, are
extremely useful for studies of AGN SEDs, especially the UV bump and
spectral break associated with the Lyman limit.
Most of our AGNs show a spectral break around
1100\AA\ (with large uncertainty for some objects), similar to what is
seen in \citet{Zhen97}, and is in agreement with the non-LTE disk
models.  The distribution of UV-optical spectral indices redward of
the break, and far-UV indices shortward of the break, are also in
rough
agreement with the models. However, we do not see a correlation
between the far-UV spectral index and the black hole mass, as predicted
by the face-on models.  Moreover, our AGNs occupy a region in
\afuv-\mbh\ space that is not covered by the thin-disk models with
\lledd\ in the range of 0.01--0.3 covered by the models.
These findings imply that some
fundamental assumptions in the models and/or in our understanding of
AGN phenomena may need rethinking.  We discuss a few relevant issues
below.

(1) We do not have information on disk inclination for our sample.
Model predictions show that large inclinations do change \afuv,
but comparing the edge-on and face-on cases, a large inclination
increases the disagreement between our data and the model
predictions in the parameter space (\afuv, \lbreak, \mbh, \lledd).

(2) Reddening can increase the observed \afuv.  We note that our data
will match the models better in \afuv-\mbh\ space if all the \afuv\
are shifted down by $\sim$1.  This will also allow us to have
non-face-on inclinations for our sample,  and it is unlikely that we
systematically under-corrected the Galactic reddening.  \citet{Scot04}
looked for possible systematic effects that could be caused by
incorrect reddening corrections in the FUSE observations of
low-redshift AGN and found none.  However, any intrinsic reddening can
play an important role.  Two objects in our sample, IRAS~F07546+3928
and PG~1351+640 show evidence of intrinsic reddening.  Both of them
show intrinsic absorption features, and unlike other objects, their
FUV to blue optical spectra significantly deviate from a power-law
(see Fig.~\ref{fg:sed}).  All the above evidence suggests a strong
(intrinsic) reddening effect.  If we assume the intrinsic reddening
has the same nature as the Galactic reddening and follows the same
extinction law \citep[][CCM]{Card89}, roughly a correction of
$\ebv=0.03$ is needed to bring \afuv\ down by about 1; for the extinction
curve of the Small Magellanic Cloud \citep[SMC,][]{Prev84}, this
requires $\ebv=0.04$.  These are modest amounts, but they have significant
effects at short wavelengths.

%\clearpage    
\begin{figure}[h]
\epsscale{.80}
\plottwo{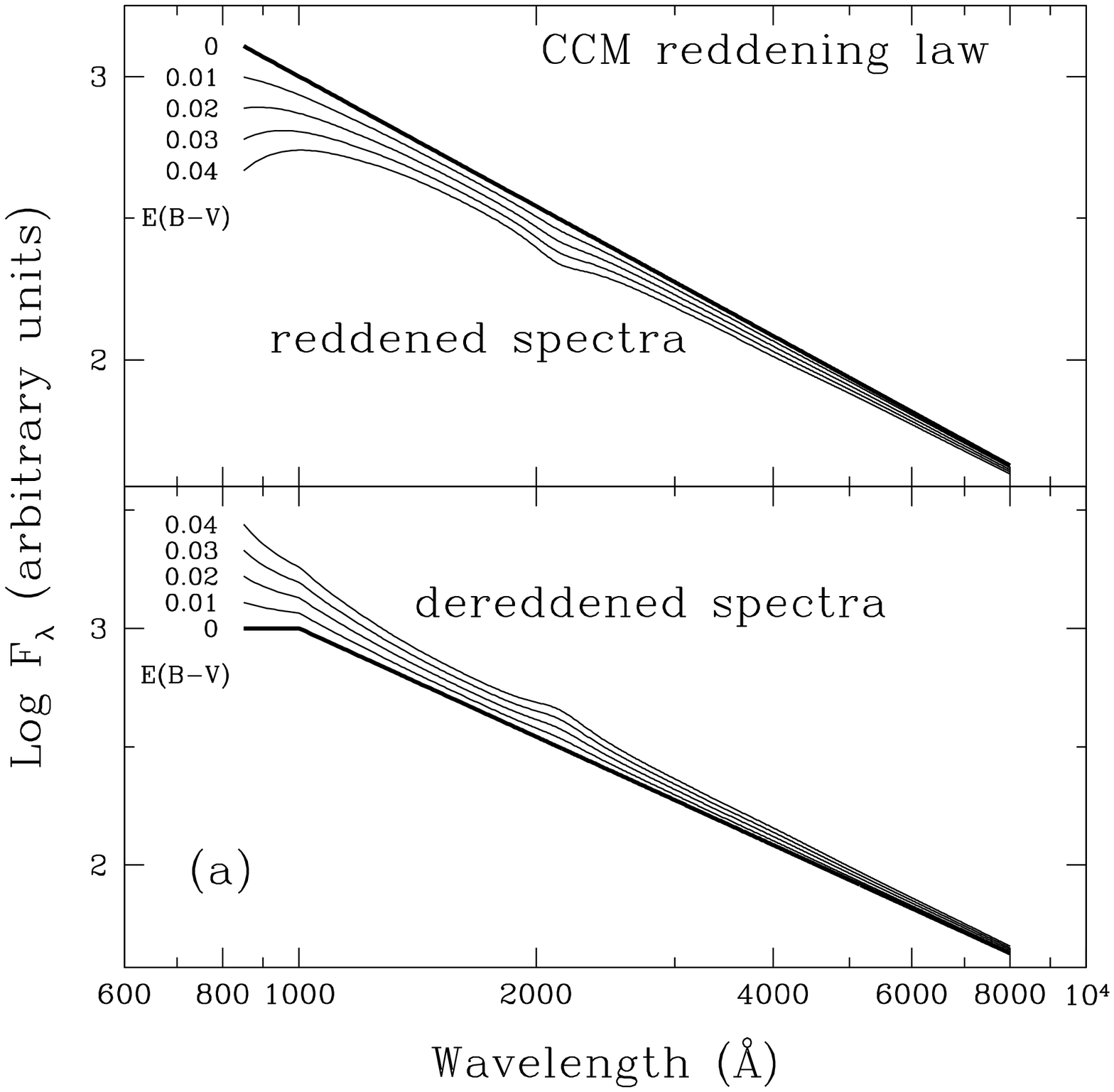}{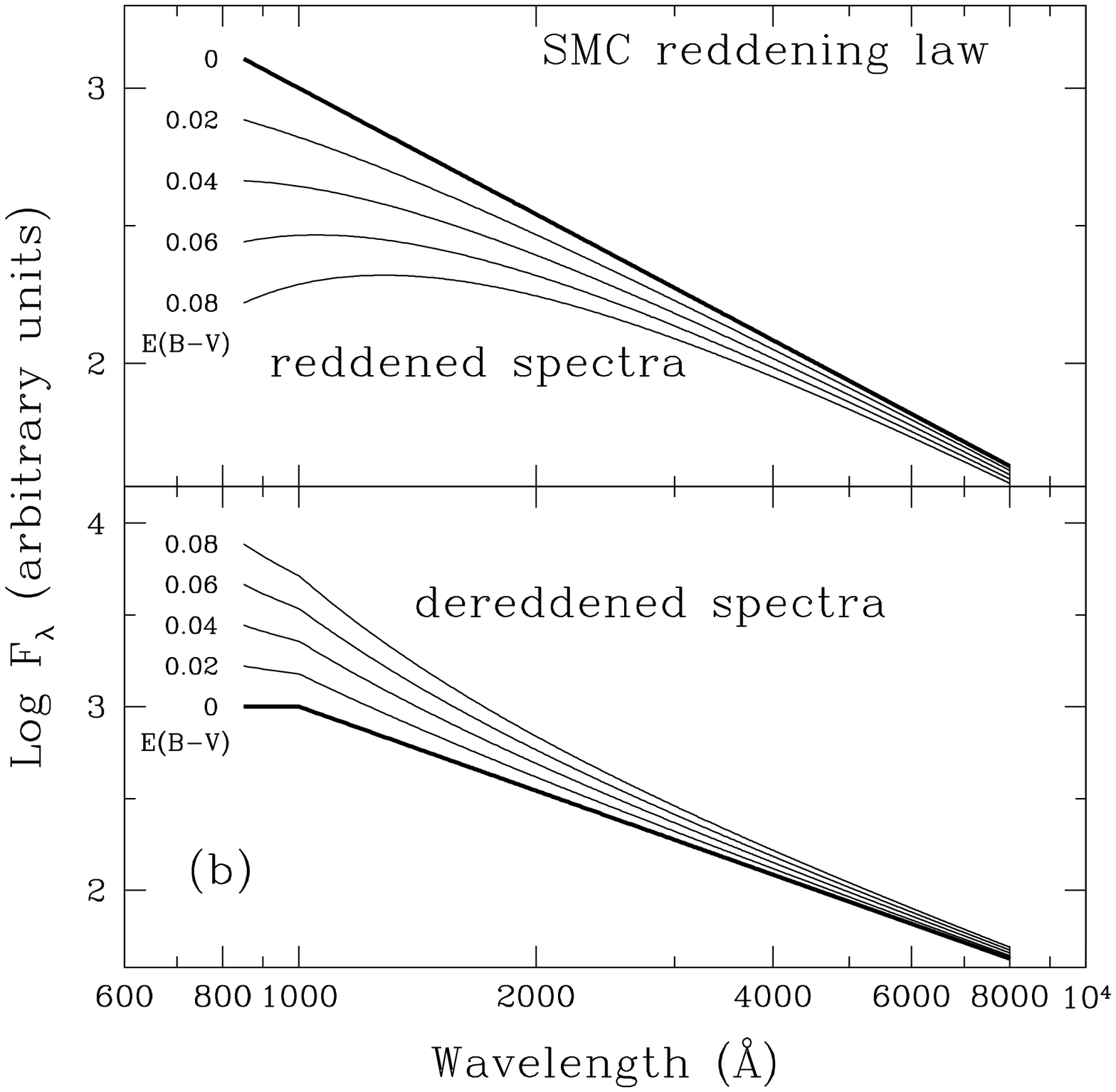}
\caption{Simulations of reddening a power-law continuum ($\auvo=-1.52$, the
median value of our sample), and dereddening a broken power-law
continuum ($\auvo=-1.52$, \afuv=0). 
(a) reddening a power-law (top, thick line) and
dereddening a broken-power (bottom, thick line) 
with the Galactic reddening curve \citep{Card89}.  
Note the spectral break at 1000\AA\ disappears (bottom)
when \ebv=0.04.
(b) same as (a) with the SMC reddening curve \citep{Prev84}. 
Note the curvature between 800--3000\AA\ (top) for \ebv=0.08.
\label{fg:simu}}
\end{figure}   %\clearpage 

To show the effect of reddening on the spectral break, we performed
some simple simulations (Fig.~\ref{fg:simu}).   We found that for
the CCM reddening curve (Fig.~\ref{fg:simu}a), adding reddening of
$\ebv=0.04$ to a power-law spectrum can produce a UV turnover that
resembles the spectral break we see in AGN spectra.  Also dereddening
a spectrum with a spectral break can virtually eliminate the break.
Compared to the intrinsic reddening seen in some nearby AGNs
\citep{Maio01}, the reddening values required here are very low
and would not be surprising to find in AGNs.  On the other hand, if we use
the flatter SMC extinction curve, a larger \ebv\ ($>0.04$) is
needed to produce noticeable results (Fig.~\ref{fg:simu}b).  However,
it cannot produce a clean UV turnover, or remove a spectral break
without introducing a large-scale curvature to the
spectrum in the NUV to optical region. \irasobj\ and PG~1351+640
in our sample seem to show this curvature (Fig~\ref{fg:sed}).

We note that the extinction curves we use above for the FUV region
are simple extrapolations from CCM and SMC extinction curves. 
\citet{HutGia01} derived FUV extinction curves for stars in the
Galaxy, Large Magellanic Cloud (LMC), and SMC using FUSE data,
and found that they appear to extend the extinction curves from longer
wavelengths in a straightforward way.
\citet{Sass02} also found an FUV extinction curve
in the Galactic diffuse interstellar medium
consistent with an extrapolation of the CCM curve ($R_V = 3.1$),
but it is likely that the Galactic extinction curve is not applicable
to AGNs, and we see significant difference between
the CCM and SMC curves in the simulations.  In fact, \citet{Maio01}
found that the ratio of \ebv\ to hydrogen column density \nhi\
in AGNs is lower than the Galactic value by a factor of $\sim
3-100$, and suggested that the dust in the circumnuclear region
of AGNs has different properties than in the Galactic diffuse
interstellar medium.  In a study of red and reddened quasars in
SDSS, \citet{Rich03} found that an SMC-like reddening law with \ebv\
between 0.135 and 0.07 can redden their normal color composites to
a dust-reddened composite spectrum.  
Normal quasars in the SDSS exhibit little to no intrinsic reddening.
\citet{Hopk04} 
find that 81\% of the SDSS quasars have $\ebv<0.02$,
and those that are reddened follow an SMC-like extinction law.
With a simple assumption that
all AGNs have the same continuum slope, \citet{Gask04} derived an
extinction curve for AGNs, and claimed that it is much flatter than
the Galactic CCM extinction curve.  Whether this is true or not,
if the reddening curve in the AGNs is at least as flat as the SMC
curve, the reddening is not able to produce the spectral break seen
in our AGNs without leaving a clear signature at longer wavelengths.  
In addition, as seen in Figure~\ref{fg:simu}b for the
SMC law, a relatively small reddening ($\ebv<0.08$) can significantly
suppress the UV continuum, as is only seen in a few of our objects.
This implies that either the intrinsic reddening for most objects
in our sample is very low, or the intrinsic reddening curve for
AGNs is very different (e.g., flatter) from the SMC curve.

It should be kept in mind that these AGNs were targeted for observation
by \fuse\ because they were known to be bright in the UV.  This
selection would bias our sample to be among the AGNs with the
least intrinsic reddening.  If our results indeed arise from this
effect, they may be stronger and more common among the general
AGN population than in our sample. 

Until reliable extinction curves and quantitative intrinsic reddening
for individual AGNs are available, reddening effects and the AGN
intrinsic continuum slope cannot be completely decoupled.  As
discussed above, intrinsic dust reddening could significantly modulate
the AGN UV bump, but it cannot produce a clear spectral break if the
extinction curve is flatter than the CCM law.  The observed spectral
break is indeed intrinsic to AGNs.

(3) Comptonization can also alter the UV and soft X-ray spectral
indices as well as smear out the Lyman edge
\citep[e.g.,][]{CzeZby91,Hube01}.  \citet{Zhen97} were able to fit
their \hst\ composite spectrum with a disk model plus Comptonization
\citep{CzeZby91}; \citet{Kris99} did the same for 3C273.  With the
same SED of 3C273, \citet{Blae01} found that while Comptonization has
no effect on the optical and mid-UV spectrum, where they got a
reasonable fit with the non-LTE disk model by \citet{Hube00},
including Comptonization is necessary to smear out the Lyman edge
feature and to extend the disk spectrum to the soft X-ray band.  It
may also be true that Comptonization is important only in some
objects, but the models we studied do not included the Comptonization
in creating the integrated spectra.  Adding Comptonization to disk
models will introduce a scattering medium and hence more parameters.
We also note that the effect of Comptonization on the model spectra is
very similar to the relativistic smearing effect which is strong in
large inclination disks.

(4) We use geometrically thin disk models to compare with our data.
The \lledd\ for many of our objects exceed the thin disk model limit
of 0.3, and there is no physical reason why \lledd\ cannot exceed this
limit.  Increasing \lledd\ would be expected to transform a thin disk
into a slim disk \citep*{Szus96}, for which advection may become
important \citep{Blae01}.  Not only does a slim disk model produce
\lledd\ that is consistent with values estimated for many quasars, it
could also improve the fit to observed spectra as suggested by
\citet{Blae01}.  Models with nonzero magnetic torques across the
innermost stable circular orbit \citep{AgoKro00} 
may also improve the fit to observed spectra.  Detailed model spectra
are needed to compare with our data.  

(5) We used models with a fixed value of \alphav\ and maximum black
hole spin to compare to our UV data.  While the FUV slope is not
sensitive to \alphav, if our objects indeed have very different black
hole spins, the predicted correlation between \afuv\ and \mbh\ for a
single value of spin may not be seen in the data. In our future work
we will construct specific models for the masses and luminosities
measured for our objects, and try to find for each object the best-fit
inclination and black hole spin.  Inclinations can also be roughly
constrained using the radio properties of radio-loud AGNs
\citep[e.g.,][]{OrrBro82}.  With fewer free parameters, the models can
be better constrained and tested by observational data.

% Accretion disk models and the physics they include are becoming more
% detailed and complex.  With well determined \mbh\ and \lledd\ from
% observations, an imminent possibility of using this data set is
% to compare the individual objects with the models by fixing these
% two parameters.  Inclinations can also be roughly constrained using
% the radio properties of radio-loud AGNs \citep[e.g.,][]{OrrBro82}.
% With fewer free parameters, the models can be better constrained
% and tested by observational data.

\section{SUMMARY\label{sec:summ}}

\begin{enumerate}
\item We construct SEDs of 17 AGNs with quasi-simultaneous spectra covering
900--9000\AA\ (rest frame).  The SEDs are available in digital format
at {\verb+http://physics.uwyo.edu/agn/+}.

\item The distribution of \auvo\ is narrow, and in rough agreement
with non-LTE thin-disk models.  The distribution of \afuv\ of our
sample is also in rough agreement with that of the models.

\item We see a spectral break in the UV for most of our objects, and
the break is around 1100\AA.  Although this result is formally
associated with large uncertainty for some objects, the FUV spectral
region is below the extrapolation of the NUV-optical slope, indicating
a spectral break around 1100\AA, in agreement with previous studies of
\hst\ composite spectra.

\item Intrinsic dust reddening
can significantly modulate the AGN continua, but the spectral break is
intrinsic to the AGNs, and is not caused by possible reddening if the
dust extinction curve in AGNs is flatter than the Galactic reddening
curve.

\item We do not find the correlation between \afuv\ and \mbh\,
expected by the thin accretion (face-on) disk model, possibly due to
the small sample size.  Scatter introduced by other varying disk
parameters that are not included in our models, such as inclination
and the black hole spin, could also weaken the expected correlation.

\item Thin-disk models do not match the observed spectra in the space
of \afuv\ and \mbh\ within the thin-disk model limit (\lledd=0.3).
This discrepancy may be attributable to the effects of 
Comptonization and other factors the models have
not included.

\end{enumerate}

% Modeling individual objects with fixed \mbh\ and \lledd\
% is now possible with this data set and can put more constraints on the
% accretion disk models.

\acknowledgments

%We are grateful to 

We thank Beverley Wills for providing the optical spectra of
PG1100+770.
This work is based on data obtained for the Guaranteed Time Team
by the NASA-CNES-CSA FUSE mission operated by the Johns Hopkins
University.  This work has been supported by NASA through grants
from the Space Telescope Science Institute, which is operated
by the Association of Universities for Research in Astronomy, Incorporated,
under NASA contracts NAS5-32985 and NAS5-26555.

%and \hst\ grant HST-AR-09913.01-A.


%% copy editing.  Individual instruments can be provided in parentheses,
%% after the keyword, but they will not be verified.

Facilities: \facility{FUSE}, \facility{HST(STIS)}, \facility{KPNO}.



\end{document}